\documentclass[12pt,a4paper,twoside,openright,arabic,english]{report}
\usepackage[sorting=none, backend = bibtex, style=numeric-comp]{biblatex}
\usepackage{float}
\usepackage[utf8]{inputenc}
\usepackage[bottom]{footmisc} 
\usepackage{xspace}
\usepackage{arabtex}
\usepackage[acronym]{glossaries}
\usepackage{textcomp}
\usepackage[table]{xcolor}
\usepackage{longtable}
\usepackage{booktabs}
\usepackage{moreverb}
\usepackage{enumitem}
\usepackage{amsfonts}
\usepackage{textcomp}								
\usepackage{lmodern}								
\usepackage{helvet}
\usepackage[english]{babel}
\usepackage[T1]{fontenc}

\usepackage{amsmath}
\usepackage{amssymb}								
\usepackage{graphicx}
\usepackage{subcaption}
\usepackage{caption}
\numberwithin{equation}{chapter}					
\numberwithin{figure}{chapter}						
\numberwithin{table}{chapter}						
\usepackage{listings}              
\usepackage{setspace}
\usepackage{color}
\definecolor{dkgreen}{rgb}{0,0.6,0}
\definecolor{gray}{rgb}{0.5,0.5,0.5}
\definecolor{red}{rgb}{0.8,0,0}
\usepackage{chemfig}								
\usepackage[top=3cm, bottom=3cm,
			inner=3cm, outer=3cm]{geometry}			
\usepackage{eso-pic}								

\usepackage{float} 									
\usepackage{parskip}								

\def\ThesisType{B} 

\definecolor{headerBrown}{RGB}{144,102,78}
\if\ThesisType B
\definecolor{thesisHeaderColor}{RGB}{126,180,56} 
\fi
\if\ThesisType M
\definecolor{thesisHeaderColor}{cmyk}{0.14,0,0,0.65} 
\fi

\usepackage[labelfont=bf, textfont=normal,
			justification=centering,
			singlelinecheck=false]{caption}

\usepackage{hyperref}								
\hypersetup{colorlinks, citecolor=black,
   		 	filecolor=black, linkcolor=black,
    		urlcolor=black}

\usepackage{titlesec}
\titlespacing{\section}{0pt}{*0}{*0}
\titlespacing{\subsection}{0pt}{*0}{*0}
\titlespacing{\subsubsection}{0pt}{*0}{*0}
\titleformat{\chapter}[display]
  {\Huge\bfseries\filcenter}
  {{\fontsize{50pt}{1em}\vspace{-4.2ex}\selectfont \textnormal{\thechapter}}}{1ex}{}[]

\usepackage{fancyhdr}								
\def\layout{2}	
\ifnum\layout=2	
	\fancypagestyle{plain}{			
	\fancyhf{}
	 		
	\fancyfoot[LE,RO]{\thepage}}	
\else			
\fi

\usepackage[textsize=tiny]{todonotes}   
\def\figurename{Fig.}

\makeglossaries
\newacronym{swap}{SWaP}{size weight and power}
\newacronym{ds}{DS}{deep space}
\newacronym{ns}{NS}{number state}
\newacronym{rf}{RF}{radio frequency}
\newacronym{fso}{FSO}{free space optical communication}
\newacronym{mm}{MM}{mode modulation}
\newacronym{pc}{PC}{photon-counting}
\newacronym{dsoc}{DSOC}{deep space optical communication}
\newacronym{nasa}{NASA}{national aeronautics and space administration}
\newacronym{dsc}{DSC}{deep space communication}
\newacronym{mlcd}{MLCD}{mars Laser Communication Demonstration}
\newacronym{llcd}{LLCD}{lunar Laser Communication Demonstration}
\newacronym{dsn}{DSN}{deep space network}
\newacronym{jpl}{JPL}{jet propulsion laboratory}
\newacronym{dsif}{DSIF}{deep Space instrumentation facility}
\newacronym{ppm}{PPM}{pulse position modulation}
\newacronym{bpsk}{BPSK}{binary phase-shift keying}
\newacronym{qpsk}{QPSK}{quadrature phase-shift keying}
\newacronym{fsk}{FSK}{frequency shift keying}
\newacronym{psk}{PSK}{phase shift keying}
\newacronym{sn}{SN}{space network}
\newacronym{maser}{MASER}{microwave amplification by stimulated emission of radiation}
\newacronym{wdma}{WDMA}{wavelength division multiple access }
\newacronym{snr}{SNR}{signal to noise ratio}
\newacronym{sep}{SEP}{sun-earth-probe}
\newacronym{spe}{SPE}{sun-probe-earth}
\newacronym{pdfs}{Pdf}{probability density function}
\newacronym{hemt}{HEMT}{high mobility electron transistor}
\newacronym{qc}{QC}{quantum communication}
\newacronym{geo}{GEO}{geostationary orbit}
\newacronym{leo}{LEO}{low Earth orbit}
\newacronym{eirp}{EIRP}{effective Isotropic radiated power}
\newacronym{dmc}{DMC}{discrete memoryless channel}
\newacronym{pdf}{PDFs}{probability density functions}
\newacronym{awgn}{AWGN}{additive white gaussian noise}
\newacronym{ber}{BER}{bit error rate}
\newacronym{tcp}{TCP}{transmission control protocol }
\newacronym{ipn}{IPN}{interplanetary network}
\newacronym{ccsds}{CCSDS}{consultative committee for space data systems}
\newacronym{sdo}{SDO}{standard-developing
organization}
\newacronym{rs}{RS}{reed-solomon}
\newacronym{scps}{SCPS}{space communications protocol specifications}
\newacronym{BSS}{bss}{bundle streaming service}
\newacronym{spps}{SPP}{spiral phase plate}
\newacronym{ip}{IP}{internet protocol}
\newacronym{dtn}{DTN}{delay/disruption tolerant networking
}
\newacronym{dstp}{DS-TP}{deep-Space transport protocol}
\newacronym{ecc}{ECC}{error correction code}
\newacronym{isecg}{ISECG}{international space exploration coordination group}
\newacronym{fom}{FoM}{figure of merit}
\newacronym{tp}{TP}{transport protocol}
\newacronym{arq}{ARQ}{automatic retransmission reQuest}

\newacronym{dttp}{DTTP}{delay tolerant transport protocol}
\newacronym{dtpc}{DTPC}{delay tolerant payload conditioning}
\newacronym{hpe}{HPE}{high photon efficiency}
\newacronym{celos}{CFLOS}{cloud free line of sight}
\newacronym{dd}{DD}{direct detection}
\newacronym{im}{IM}{intensity modulation}
\newacronym{scppm}{SCPPM}{serially concatenated pulse position modulation}
\newacronym{lgb}{LGB}{laguerre-gaussian beam}
\newacronym{sam}{SAM}{spin angular momentum}
\newacronym{oam}{OAM}{orbital angular momentum}
\newacronym{at}{AT}{atmospheric turbulence}
\newacronym{fec}{FEC}{advanced forward error correction}
\newacronym{sq}{SQ}{single-quadrature}
\newacronym{edfa}{EDFA}{erbium-doped fibre amplifier}
\newacronym{ppb}{PPB}{photon per bit}
\newacronym{tdrss}{TDRSS}{tracking and data relay satellite system}
\newacronym{qkd}{QKD}{quantum key distribution}
\newacronym{qt}{QT}{quantum teleportation}
\newacronym{qss}{QSS}{quantum secret sharing}
\newacronym{fer}{FER}{frame error rate}
\newacronym{qsdc}{QSDC}{quantum secure direct communication}
\newacronym{ldpc}{LDPC}{low density parity check}
\newacronym{ook}{OOK}{on-off keying}
\newacronym{apd}{APD}{avalanche photo diode}

\newcommand\Ou{%
  \mathrel{{\ooalign{\hss\raisebox{-0.5ex}{$-$}\hss\cr\raisebox{0.5ex}{$+$}}}}}

\begin{document} 



\ifx\ThesisType\undefined
Undefined Thesis type in settings.tex
\else
  \if\ThesisType M
    \else
    \if\ThesisType B
    \else
    Define \textbackslash ThesisType as M for master's thesis or B for bachelor thesis in settings.tex
    \fi
  \fi
\fi

\begin{titlepage}
\newgeometry{top=3cm, bottom=1cm,
			left=2.25 cm, right=2.25cm}	
			
\ifx\ThesisType\undefined
\else
    \if\ThesisType M
    \vtop{
        \null\vspace{-25mm}
        \centerline{\includegraphics[width=1.18\textwidth]{}}
        \vspace{-2.3cm}
        \hbox{\hspace{0mm}\includegraphics[height=18mm]{}}
        \centerline{\textcolor{headerBrown}{\rule{1.18\textwidth}{4pt}}}
        \vspace{\paperheight}\vspace{-85mm}
        \centerline{\textcolor{thesisHeaderColor}{\rule{1.1\textwidth}{0.8pt}}} 
        \vspace{-\paperheight}\vspace{85mm}
    }
    \fi
    \if\ThesisType B
    \vtop{
        \null\vspace{-25mm}
        
       \centerline{\includegraphics[width=1\textwidth,height=85pt]{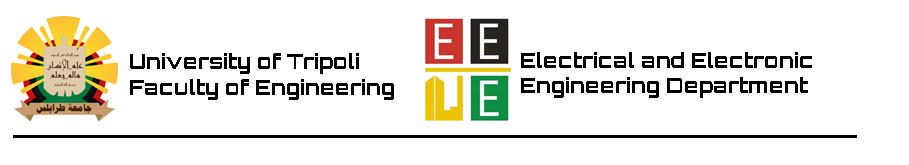}}
        \vspace{-2.3cm}
        \vspace{\paperheight}\vspace{-85mm}
      
        \vspace{-\paperheight}\vspace{85mm}
    }
    \fi
\fi

\vspace{1cm}
\begin{figure}[H]
\centering
\vspace{3cm}	

\includegraphics[width=0.7\linewidth]{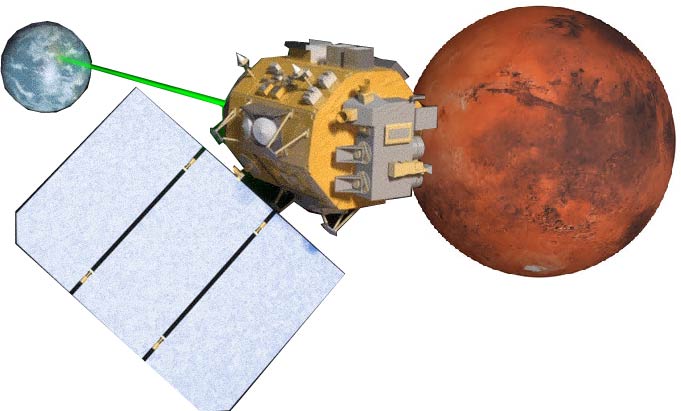}

\end{figure}
\vspace{1cm}
\renewcommand{\familydefault}{\sfdefault} \normalfont 
\vspace{1cm}
\centering
\textbf{{\Huge  Connecting the Universe: Challenges, Mitigation, Advances, and Link Engineering }}\\[0.3cm]
\textbf{{\Huge   }}\\[1cm]
\vspace{1cm}
{\Large A Project Report Submitted in Partial Fulfilment of the Requirements for the Bachelor of Science Degree in  Electronics and Communication Engineering}\\[1cm]
\text{Prepared by: Sara Mahmoud Karmous} \\
\text{Supervised by: Dr. Nadia Adem} \\
\text{Fall 2019 } \\
\text{Tripoli - Libya} \\

\end{titlepage}

\newpage
\restoregeometry
\thispagestyle{empty}
\mbox{}


\newpage
\pagenumbering{roman}	

\thispagestyle{plain}			
\setlength{\parskip}{0pt plus 1.0pt}

\section*{Abstract}

With the large number of  deep space (DS) missions anticipated by the end of this decade,  reliable-high capacity  DS communication systems are needed more than ever. 
Nevertheless, existing DS communication technologies are way far from meeting such a goal.  
Improving current systems does not only demand a system engineering leadership, but more crucially a well investigation in the potentials of  emerging technologies  in overcoming the challenges of the unique-ultra long DS communication channel. This project starts with a survey, with the fact that there has been no single one on DS communication  published in the literature over the last decade. Such  surveys are not only essential to demystify the field and overcome  difficulties arising as a result of its overlap with so many other fields. They are as well lacking to highlight current technologies, trends, and advancements, investigate potentials, identify challenges, and in essence provide perspectives and propose solutions. 
The survey is comprehensive that indeed covers all the aforementioned aspects.
The survey (presented mainly in the corresponding chapters), focus on free space optical (\acrshort{fso}) communication as a potential technology that can overcome the shortcomings of current radio frequency \acrshort{rf} based communication systems.
For the best of our knowledge, in addition it provides for the  very first time a thoughtful discussion about implementing orbital angular momentum (\acrshort{oam}) for DS and  identifies major related challenges and proposes some novel solutions.
Furthermore, we  discuss  DS modulations, and coding schemes, as well as DS emerging receiver technologies and communication protocols.
We also elaborate on how all of these technologies  guarantee reliability, improve efficiency,  offer  capacity boost, and enhance security in the unique \acrshort{ds} environment. In addition to that an extended study on design and performance analysis of deep space optical communication \acrshort{dsoc} is included, with the most suggested modulation for such a link which is pulse position modulation (\acrshort{ppm}), and a focus on the communication between Earth and planet Mars which is an important destination for space exploration. All graphs, unless otherwise cited,  are generated using  MATLAB, and diagrams are designed using  PhotoShop and Illustrator. In addition, all the survey tables are originally designed and created. \\

\textbf{keywords}: Avalanche photo-diode (\acrshort{apd}) detector, deep space (DS) communications, free space optical (FSO) communications,   Holevo limit, interplanetary network (\acrshort{ipn}), orbital angular momentum (OAM), photon-counting  (\acrshort{pc}) detector, pulse position modulation (PPM),  quantum communications (\acrshort{qc}). mode modulation (\acrshort{mm}). 

\newpage

\let\savecleardoublepage\cleardoublepage
\let\cleardoublepage\clearpage 
\thispagestyle{plain}			
\section*{Acknowledgements}

 I want to thank my supervisor Dr. Nadia Adem for her guidance throughout the project. I would like also to thank our research partner Dr. Bassem Khalfi for his contributions in working through the survey. I take this opportunity to express gratitude to my family, friends and to all of the department faculty members for their help and support.

\newpage

\begin{otherlanguage}{arabic}

\textbf{الملخص}\\

مع العدد الكبير من مهام الفضاء العميق \textLR{ deep space (DS)} المتوقعة بنهاية هذا
العقد، هناك حاجة إلى أنظمة اتصالات \textLR{DS} موثوقة ذات سعة عالية أكثر من
مضي. إن تقنيات اتصالات \textLR{DS} الحالية بعيدة كل البعد عن تلبية احتباجات مثل هذا القناء. لا يتطلب تحسين الأنظمة الحالية فقط لقيادة هندسية للنظام، ولكن الأهم من ذلك هو إجراء تحقيق في إمكانات 
التقنيات الحديثة في التغلب على تحديات اتصالات \textLR{DS} .الفريدة التي تعمل علي مسافات طويل جدًا
.يبدأ هذا المشروع بدراسة استقصائية، مع علم أنه يتم نشر ورقة علمية استقصائية 
   في مجال اتصالات  الفضاء \textLR{DS} على مدار العقد الماضي.
هذه الاستطلاعات ليست ضرورية فقط لإزالة الغموض عن المجال و 
 التغلب على الصعوبات الناشئة نتيجة تداخلها مع العديد من المجالات الأخرى. كما أنها تفتقر أيضًا إلى إبراز التقنيات والاتجاهات والتطورات الحالية، والتحقيق في الإمكانات، وتحديد التحديات، وفي جوهرها تقديم وجهات النظر واقتراح الحلول. الدراسة شامل وتغطي بالفعل جميع الجوانب المذكورة أعلاه. تركز هذه الدراسة على الاتصال البصري للفضاء الحر
\textLR{free space optical communications (FSO)}باعتباره تقنية يمكنها التغلب على اغلب أوجه القصور في أنظمة الاتصالات القائمة على الترددات الراديوية الحالية. وبالإضافة إلى ذلك، نقدم لي اول مرة مناقشة حول تطبيق الزخم الزاوي المداري \textLR{orbital angular momentum (OAM)} لـ \textLR{DS} وتحديد التحديات الرئيسية ذات الصلة واقترح بعض الحلول الجديدة. وبالاضافة الي ذلك، نناقش
 تشكيلات  شباكات \textLR{DS} وطرق التشفير، فضلاً عن تقنيات مستقبلات \textLR{DS} الحديثة
وبروتوكولات الاتصال. نوضح أيضًا كيفية استخدام كل هذه التقنيات
وضمان الموثوقية وتحسين الكفاءة و تعزيز السعة والأمان
في بيئة \textLR{DS} الفريدة. بالإضافة إلى ذلك، قمنا بدراسة موسعة حول تصميم وتحليل أداء الاتصالات البصرية في الفضاء DSOC ،\textLR{deep space optical commutations (DSOC)} ،
مع أكثر الأنظمة التضمين المقترحة لمثل هذا النوع من الاتصالات هو تضمين موضع النبضة
 \textLR{pule position modulation (PPM)} ، والتركيز على التواصل بين الأرض وكوكب المريخوالتركيز على الاتصال بين الأرض و كوكب المريخ علي انها وجهة مهمة لاستكشاف الفضاء. يتم إنشاء جميع الرسوم البيانية باستخدام  \textLR{MATLAB} ،وتم تصميم الرسوم البيانية باستخدام \textLR{PhotoShop}
 \\
 \\
 \\
 الكلمات الرئيسية: كاشف الانهيار الضوئي للصمام الثنائي \textLR{(APD)} ،اتصالات الفضاء السحيق \textLR{DS} ،الاتصالات البصرية الهوائية الحرة \textLR{(FSO)}  ،حد هوليفو ، شبكة  بين الكواكب \textLR{(IPN)}الزخم، الزاوي المداري \textLR{(OAM)} ،عداد الفوتون \textLR{(PC)} كاشف، تعديل موضع النبض \textLR{(PPM)} ،اتصالات الكم \textLR{(QC)}. تضمين الحالة \textLR{(MM)}.

\end{otherlanguage}

\tableofcontents
\printglossary[type=\acronymtype]

\cleardoublepage
\addcontentsline{toc}{chapter}{\listfigurename} 
\listoffigures

\addcontentsline{toc}{chapter}{\listtablename}  
\listoftables

\cleardoublepage
\setcounter{page}{1}
\pagenumbering{arabic}			
\setlength{\parskip}{0pt plus 1pt}

\newcommand{\red}{\textcolor{red} }
\chapter{Introduction}
\thispagestyle{plain}			
\setlength{\parskip}{0pt plus 1.0pt}
\renewcommand{\figurename}{Fig.}
 \captionsetup[figure]{labelformat=parens, labelsep=colon,}

Space exploration not only offers an opportunity to understand the universe, but also results in endless benefits for diverse aspects of human's life such as technologies, computing, transportation, environment, economy, to name a few~\cite{ISCEG2013}.
Therefore, there is an increased investment on global exploration from a number of worldwide space agencies~\cite{NASA2021plan} as well as private companies~\cite{SpaceXMoon}, which all are aiming for expanding human presence and knowledge into the Solar system~\cite{edwards2017update,ISECG2018roadmap}.
For instance, in addition to the number of recent-successful missions to Mars (e.g.~\cite{CNSA2021Tianwen,ProvencesMars2020}), there are some other missions scheduled for the next few years.  ExoMars and Tera-hertz are  two of the major scheduled missions~\cite{Exomarsprogram2022,larsson2017mars}. While the goal of the former is to  search of live in Mar,  the purpose of Tera-hertz explorer is to investigate the chemical process to understand Mars atmosphere. The very-first Mars crew mission is also scheduled for the   2024-2026 timeline~\cite{FirsthumanlandingMars2026}. 

\section{Motivation} 
One of the major drivers for the success of space exploration, however, is facilitating broadband interplanetary communication~\cite{ESA20Mars}. 
 Yet, reliable interplanetary networks will only be possible through the development of technologies and protocols that are suitable for facing the unique \acrshort{ds} communication challenges. Unlike Earth and near-Earth communication, DS suffers from  non-guaranteed line-of-sights, ultra-long time-changing distances, variability of Sun-Earth-probe angles, solar scintillation, and atmospheric conditions which all hinder communication, tremendously deteriorate  quality, and hence limit capacity~\cite{seas2019optical}.  
Current missions to Mars, for example, have data rates of only few megabits per second at the minimum Earth-Mars distance~\cite{arapoglou2017benchmarking,kwok2009DSNHandbook}. 
The capacity  drops order of magnitude as the distance increases~\cite{AmalTechSpec2021}.
    Things become even worse for beyond Mars  missions. The Voyage probe located at 22 billion Km distance, for example, operates at the maximum of few Kbits per second~\cite{JplVoyage}. 

\begin{figure}[t]
\centering
\includegraphics[width=1.0\linewidth]{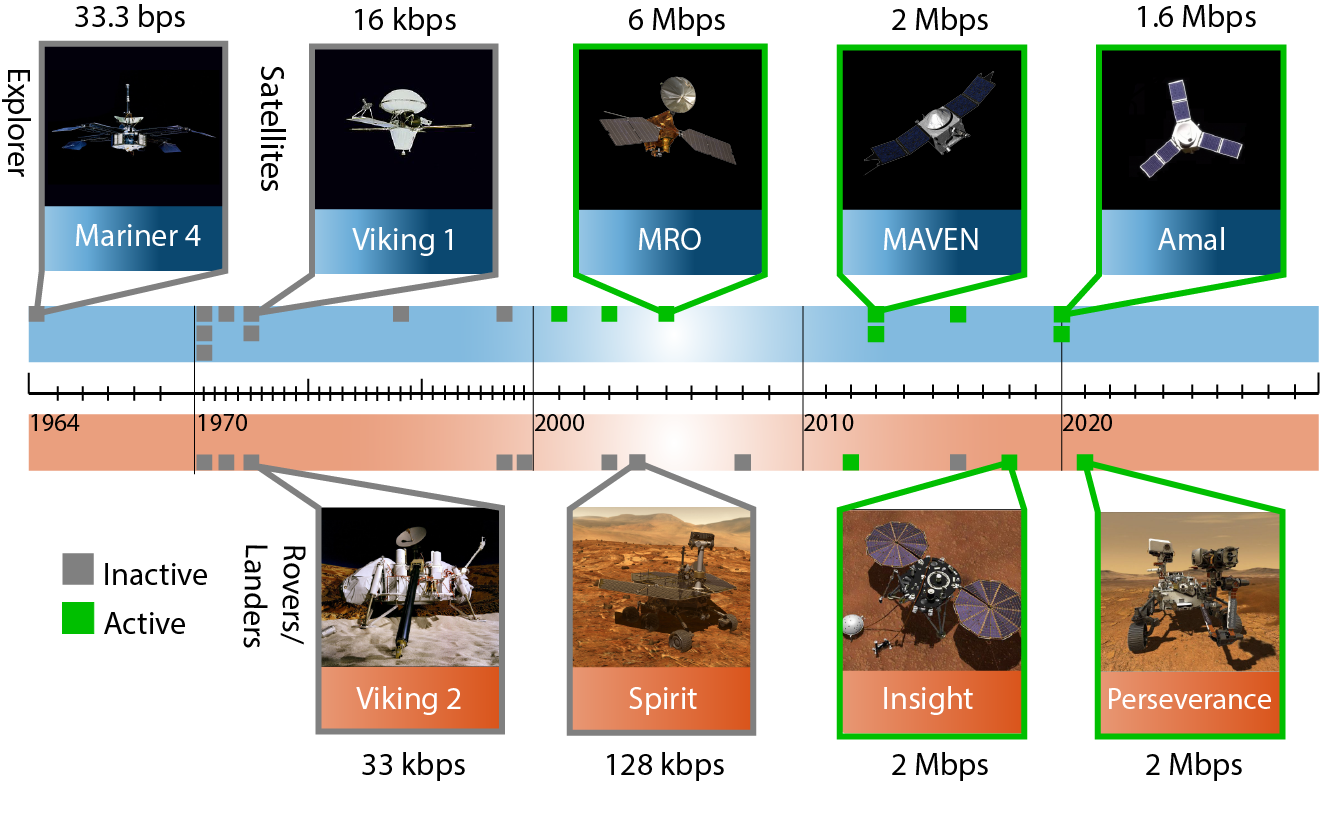}
\caption{Mars missions timeline.}
\label{Fig.Timeline}
\end{figure}


 Fig.~(\ref{Fig.Timeline}) illustrates the timeline for Mars orbiters, landers, and rovers and some corresponding data rates~\cite{AmalTechSpec2021,Russianspaceweb,MissiontoMars,ListofartificialMars,ListofMarsorbiters,JPL2021Mars,maevan101,NasaViking2,taylor2016mars,David2020Marina4,NasaViking1Orbiter,ProvencesMars2020}. The timeline also includes the very first Mars explorer, namely  Marina 4  which captured the first close up images for Mars and transmitted them to Earth  at the maximum of 33.3 bps data rate~\cite{David2020Marina4}.

To build  high-capacity DS communication links, research efforts have been devoted to explore the potential of \acrfull{fso} communication~\cite{lyras2019deep,seas2019optical,hemmati2011deep,kaushal2016optical}.
Compared to \acrshort{rf}-based counterparts, \acrshort{fso} communications enjoy larger bandwidth, smaller antennas at final stations, much improved security, less transmit power, and higher immunity against interference.
Nevertheless, tremendous work is yet to be done to make the FSO technology more mature and ready to be implemented~\cite{fielhauer2012concurrent}. In this regard, some projects have been adopting FSO  demonstrations for near-Earth, Moon, and Mars  communication with  plans to reach even Jupiter~\cite{grein2015optical,wilson1997results,biswas2006mars,cesarone2011deep,oh2017development,biswas2018deep}. 

The success of  adopting laser \acrshort{ds} communications will play an important role in  implementing broadband interplanetary networks. Achieving such a goal, though, will in addition be determined by the reliability, effectiveness, as well as efficiency of so many other technologies among which are coding, modulation, detection, and networking protocols.  
 The criteria for  selecting DS communication schemes is tremendously different than that for terrestrial,  airborne, and even space communication. pointing out the various DS communication system design criteria. In addition, presenting the  latest research and innovations,  propose new ideas, and highlight new directions  in meeting them. 
\section{Contributions}
The goal of work is to emphasize on how technologies are shaping   and potentially advancing \acrshort{ds} communication. 
  The level is comprehensive such that readers can grasp in one place, the overall idea about  existing DS communication systems, their limitations, the need for  improving them and their challenges. Communication technologies are covered from the prospective of DS, namely we study how they shape the DS communication and the issues resulting specifically from implementing them in the DS environments. 
  With that goal in mind, we aim, as a consequence, to   increase awareness of those challenges and hence facilitate more research progress in the field.
Readers  are to be provided with a comprehensive set of references that may address any specific aspect of the field in more details. 
  The need for providing our work is mainly driven by  $i)$ the lack of recent surveys covering this topic, and $ii)$ the existence of surveys that only cover certain-specific technologies adopted by DS communication and, in most cases, not even in the context of DS in particular. 
In Table \ref{Table:RelatedSurvComp}, comparing this work with the other related  surveys.

The contributions of this survey are fivefold, summarized as follows.
\begin{enumerate}[label=\arabic*)]
\item We survey major DS communication challenges and the state-of-art addressing  them.  
\item We cover the most-widely used modulation and coding schemes  and the opportunities of enhancing DS communication performance through fusing them with some other emerging technologies like \acrshort{fso}, detection techniques, and \acrfull{oam}.  
\item  We vulgarize new promising technologies that overlap between communications and physics. The lack of doing so seems to be making communication society shy away from contributing to the development of many related techniques. 
\item We present an overview of the current implemented \acrshort{ds} communication system, study existing visions for its future and outline the challenges in implementing them.
\item To the best of our knowledge, for the  very first time, we provide  thoughtful discussions about implementing OAM for DS and  identify major related challenges and propose some novel solutions. In terrestrial communications, OAM has been receiving a growing attention  in meeting the ever increasing high capacity demand~\cite{trichili2019communicating}, yet, only sporadically, one can find a work addresses the potentials of \acrshort{oam} for the DS.
\end{enumerate}
 With our survey, we aim in paving the way for  more rapid  advancements  in DS communication technologies and protocols 
to  empower  universe exploration and accessibility missions with   robust communications services.

\begin{center}
 
\begin{table*}[htbp]
\centering 
 \caption{Deep space communication surveys comparison}
\label{Table:RelatedSurvComp}
\resizebox{\textwidth}{!}{\begin{tabular}{|p{2.2cm}|p{2.5cm}|c|c|c|c|c|c|c|c|}
\hline
Classification              & Main-content          & Hemmati~\cite{hemmati2011deep} & Cesarone~\cite{cesarone2011deep} &Tomaso~\cite{de2011reliability}&  Zhang~\cite{zhang2011survey} & Kaushal~\cite{kaushal2016optical}  & Wu~\cite{wu2018overview} & Biswas~\cite{biswas2018deep} &This work \\
\hline
Overview     & Missions timeline      & $\times$     & $\times$   & $\times$     & $\times$        & $\times$ &$\surd$      &$\surd$       & $\surd$       \\  \cline{2-10}
             & Demonstration         &$\surd$         &$\surd$       &$\times$      & $\times$         &$\surd$    &$\surd$    &$\surd$         &$\surd$  \\ \cline{2-10}
             & Exiting \mbox{system}         & $\times$   & $\surd$      & $\times$    & $\times$       &  $\times$   &$\surd$    &$\surd$         & $\surd$     \\ \cline{2-10}
             & RF system              & $\times$  & $\surd$       & $\times$    & $\surd$   &  $\times$   &$\times$     &$\times$          & $\times$     \\ \cline{2-10}
             & FSO                     & $\surd$   & $\surd$      & $\surd$    &$\times$   &  $\surd$    &$\surd$    &$\surd$         & $\surd$     \\ \cline{2-10}
            & Goals~and limitation     & $\surd$     & $\surd$    & $\surd$     & $\surd$        & $\surd$   &$\surd$     &$\surd$    & $\surd$      \\ \cline{2-10}
            & Development details       & $\times$    & $\times$     & $\surd$     & $\times$        & $\times$   &$\times$   &$\surd$       & $\times$    \\ \cline{2-10}
             & Future prospective      & $\surd$      & $\surd$   & $\surd$     & $\surd$        & $\times$ &$\surd$     &$\surd$     & $\surd$      \\ \hline
Basics       & Link \mbox{engineering}        & $\surd$     & $\times$   & $\times$    & $\surd$        & $\times$  &$\times$    &$\times$       & $\surd$   \\ \cline{2-10}
             & Receiver \mbox{technologies}   & $\surd$      & $\surd$    & $\times$     & $\surd$            & $\surd$     &$\times$        &$\surd$       & $\surd$   \\ \cline{2-10}
             & Overall \mbox{challenge}      & $\surd$      &  $\surd$      &  $\times$    & $\surd$    & $\times$   &$\times$      &$\times$          & $\surd$   \\ \cline{2-10}
              & Turbulence effect      & $\surd$      & $\times$    & $\times$    & $\times$       & $\times$   &$\times$     &$\times$          & $\surd$         \\ \hline
Enabling technologies  & Protocols      &$\times$    & $\times$   & $\surd$      & $\surd$     & $\surd$ &$\times$        &$\times$          & $\surd$         \\ \cline{2-10}
                  & Modulation          & $\surd$      & $\surd$   & $\times$       & $\surd$     & $\surd$  &$\times$      &$\times$          & $\surd$         \\ \hline
                 &  Coding              & $\surd$       & $\surd$    & $\surd$   & $\surd$      & $\surd$    &$\times$   &$\times$           & $\surd$        \\ \hline
Performance analysis  & Capacity        & $\surd$     & $\surd$      & $\times$     & $\times$       & $\times$  &$\times$    &$\times$           & $\surd$       \\ \hline
Advances            & OAM               & $\times$    & $\times$     & $\times$       & $\times$        & $\surd$  &$\times$    &$\times$         & $\surd$        \\ \cline{2-10}
                   & IPN              &$\times$        & $\times$   & $\times$          & $\surd$         & $\times$ &$\times$     &$\times$          & $\surd$         \\ \cline{2-10}
                   & Quantum         & $\times$        & $\times$    & $\times$       & $\times$      & $\times$   &$\times$    &$\times$           & $\surd$  \\ \hline
\multicolumn{2}{|c|}{Time relevant}             & 2011   & 2011   & 2011   & 2011       & 2017    & 2018       & 2018     & 2021       \\ \hline
\multicolumn{2}{|c|}{DS context }              & Fully   & Fully    & Fully & Fully  & Partially    & Fully    & Fully   & Fully  \\ \hline

\end{tabular}}
\end{table*}
\end{center}

\section{Structure}
This thesis is structured as follows. After starting with the introduction which included the motivation and existing system with its raised issues,  chapter~\ref{chp:dscom} addresses the challenges   and proposing the solution, with the  advantages of using \acrshort{fso} communication for better performance system in section~\ref{sec:dscomb}. After that in subsection~\ref{subsec:dscomPPM} analysing the most suggested modulation for DSOC which is \acrfull{ppm}, also with types of coding used and their performance. Section~\ref{sec:dscomRec} discussing different receivers technologies and challenges with its corresponding modulation, including the photon-counting receiver. Also capacity limit performance for different modulations and suggestions for how to reach the ultimate capacity, with results for data rates in different planet's distances included. In section~\ref{sec:dscomSP}, investigation of the type of protocols used in space network and what aspects need to be improved are discussed. In chapter~\ref{ch:dsAdvances} starting with the huge capacity return by using OAM in section~\ref{OAM} and discussing how it could be deployed in DS with its challenges. Next addressing \acrfull{ipn} advances. Then in section~\ref{Quanta} an investigation on the possibility of using quantum communication in space is discussed. In chapter~\ref{lastchapter}, a design study and performance analysis for using \acrshort{ppm} with photon-counting detectors are presented for DS link in Mars range and beyond.


\chapter{Challenges and Enabling Technologies}
\label{chp:dscom}
\thispagestyle{plain}			
\setlength{\parskip}{0pt plus 1.0pt}
\renewcommand{\figurename}{Fig.}
 \captionsetup[figure]{labelformat=parens, labelsep=colon}

\section{DS Optical Communication}
\label{sec:dscomb}
Deep space optical communication (\acrshort{dsoc}) promises tens to hundredfold increase in the data rates, thanks to the high available bandwidth in the licence-free spectrum in which the technology  operates~\cite{kaushal2016optical,khalighi2014survey}. 
Such a capacity improvement has already been proven in space communication and partly for some local-links in the deep space. 
The gain of using \acrshort{fso} will have more substantial improvements when it comes to implement in \acrshort{ds} communication distances. For example, compared to \acrfull{geo} satellites, a $60$ to $80dB$ gain is required to reach Mars~\cite{hemmati2011deep}.
Such a high required gain can be achieved through some of the following:

\begin{enumerate}[label=\alph*.]
    \item Increasing the power efficiency by relying on laser beams and large transmitters gain,
    \item Accurately tracking the space terminals using beacons to point lasers towards them prior to sending data,
    \item Relying on large gain receivers (large antenna diameters),
    \item Considering advanced signaling and detection techniques that maximize  photon efficiency (number of bits per photon). 
\end{enumerate}

\begin{figure}[h]
\centering
\includegraphics[width=0.75\linewidth]{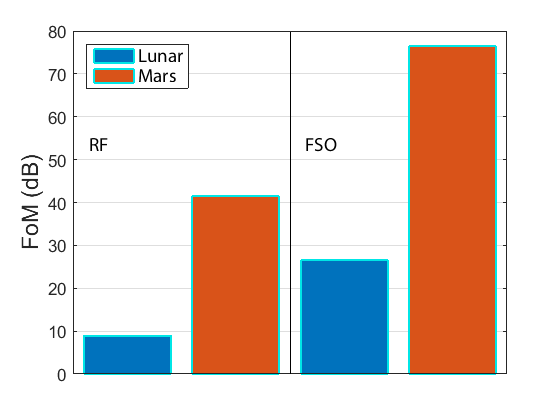}
\caption{Future communication systems FoM.}
\label{Fig.ComRfFSO}
\end{figure}

To determine the gain  of using \acrshort{fso} compared to RF communication, we define,  the \acrfull{fom} in ($dB$) as 
\begin{equation}
{10}\times{\frac{D_{is}^{2}\times{R}}{{D_{ia}}\times{P}}},
\end{equation}
where $Dis$ in the DS link distance in astronomical unit (AU), one AU is 149,597,870.7 km, which is the distance from Earth to the Sun, its used to represent huge distances in the million kilometers range. $R$, $P$, and $Dia$ are the data rate in bits per second, communication power in Watts, and receiver diameter size in meters, respectively~\cite{fielhauer2012concurrent}. 
As it counts for  space terminal performance (data rate) and cost (receiver aperture size and communication power) at any distance, the FoM metric provides a direct comparison to both the RF and FSO technologies at the system level.
Based on  future projection available in~\cite{arapoglou2017benchmarking},\cite{wu2018overview} and~\cite{cornwell2017nasa}. Fig.~(\ref{Fig.ComRfFSO}), represents the \acrshort{fom} for Lunar and Mars both for \acrshort{rf} and \acrshort{fso} systems. It should not come as a surprise that an FSO system represents more than $30dB$ advantage over the RF counterpart for Mars, while just less then $20dB$ is obtained for Lunar.

Despite the advantages that \acrshort{fso} offers, a careful consideration of the near-Sun effect on the signal strength is needed in order to compensate for that without harming the receiver circuits. 
Some techniques have already been proven to provide an increased photon efficiency such as the use of \acrfull{ppm} with \acrfull{pc} detectors. This receiver technology can provide up to $20 dB$ gain over other techniques such as \acrfull{apd} or direct detection optical pre-amplifiers~\cite{hemmati2011deep}. We will cover these techniques in more detail in  subsequent sections.

\section{RF  in Future DS Communication}
In the rest of this chapter and the next, we  will mainly focus on optical communication challenges and mitigation and advances as being a very promising technology for future DS communication. Nevertheless, since there is some interest in having a hybrid system implemented in future deep space communication~\cite{gueye2020performance}, it is of importance to cover the  main aspects of the RF technologies as well. The main reasons for the inclusion of   \acrshort{rf} in the Earth-to-probe link communication are $i)$ cost saving  and technology maturity   $ii)$ link challenge variations. 

\section{DSOC Challenges}
\label{sec:dscom}
In this section, we discuss the different challenges facing DS optical communication especially at the link and  protocol levels. We also address some of promising technologies to overcome them and the issues facing implementing these technologies.

\label{sec:dscoma}
Ensuring high capacity and reliable \acrshort{ds} communications faces different challenges:
\begin{enumerate}[label=\alph*.]
\item The distance between Earth and probes  (satellites in other planet orbiter, rovers, etc) is ultra long and time changing. 
\item Interference from other missions,  Sun,  other celestial bodies, and cloud cover  that can impact link availability.  
\item Solar  scintillation, atmospheric  conditions, and other channel impairments.
\item Variability of  Sun-Earth-probe angles and the effect of that on blocking  line-of-sight links for quite some time. 
\item The concerns from mission designers on the burden   of size, \acrfull{swap} of spacecraft communication systems. 
\item The readiness and availability of transmit and receive technologies and communication protocols that meet desired performance in the presence of the aforementioned challenges and concerns. 
\end{enumerate}
First, the communications with targets in the \acrshort{ds} requires sending signal over long distances, leading, as a consequence, to strong power dissipation and hence signal attenuation.
Even with the use of relatively large gain transmit and receive antennas, the received power can still be  very weak below the detection sensitivity.  
Also, the long distance results in a inevitable-large latency that prevents real-time communication and control.
Two-way  communication with  Mars, for example, may undergo 22 minutes transmission delay~\cite{Mars3Missions2021}.
Furthermore, the Solar flares are shown to affect the communication signals making them weaker and noisier~\cite{hemmati2011deep}. In addition, solar scintillation and turbulence affect the communication.
Also, due to the movement of Earth and targets as well as the relays in their orbits, the link capacity gets hugely affected. The Sun may completely block a communication path. 
Lastly, existing technologies at all levels especially at physical and protocol cannot be adopted directly for DS. New technologies and advanced techniques are needed to overcome all aforementioned issues and satisfy any SWaP constraints.

A potential candidate to enhance the link performance is the use of free space optical communication. 

\subsection{Modulation and Coding for DS Communication}
\label{subsec:dscomPPM}

\begin{figure}[H]
\centering
\includegraphics[width=0.9\linewidth]{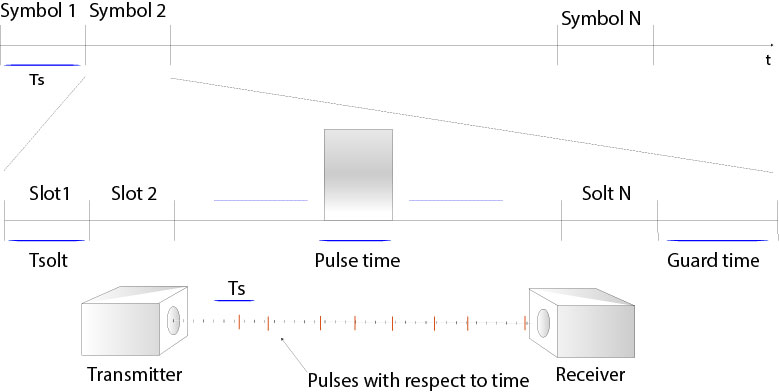}
\caption{ PPM modulation }
\label{PPM}
\end{figure}

Conventional RF-based space communication rely commonly on \acrfull{bpsk} modulation scheme~\cite{statman2004coding}. \acrshort{bpsk} is favored thanks to the low bit error rate in the low signal to noise ratio regime.
There are  common modulations schemes for \acrshort{dsoc}, which are direct detection formats, \acrfull{ook}, \acrfull{ppm}, wavelength modulation, and pulse intensity modulation.
Among the aforementioned schemes,\acrshort{ook} and \acrshort{ppm} are the most recommended.
Even though they are similar in concept, PPM is more preferred 
than OOK for DSOC as the former is being more power efficient. On the other hand, OOK is more suitable for near-Earth links~\cite{hemmati2011deep}.
OOK modulates one as a pulse and  zero with the absence of signal. \acrshort{ppm}, however, which is often used with PC detector~\cite{hemmati2011deep,hemmati2006deep,li2012dual}, is a pulse time modulation technique as shown in Fig. (\ref{PPM}), works by  dividing each channel symbol period into $M$
time slots. $k_{b} = log_{2}M $ bits are modulated by transmitting a single pulse in one of the $M$ possible time slots  $T_{slot}$ of a symbol time $T_{s}$, see Eq. (\ref{M}),(\ref{Ts}) and (\ref{Tslot}).
In \acrshort{ppm}, the information is encoded in the position of the pulse, making it robust against noise, but requires high bandwidth~\cite{moision2005coded}. The good news, though, is that high bandwidth is readily offered by \acrshort{fso} and hence is not of much concern.

To characterize the effectiveness of PPM for DS applications, we discuss the  capacity offered by PPM-\acrshort{pc} receiver system in the presence of noise and using Poisson channel model (typically used for DS system characterization~\cite{kakarla2020one}). The approximate system capacity, denoted by $C_{PCR}$, is given as~\cite{tyson1996adaptive}:

\begin{equation}
C_{PCR}= \frac{1}{\mathrm{{ln}(2) }{E}} \Bigl(\frac{{P_r}^2}{P_n\frac{2}{M-1}+\frac{P_r}{\mathrm{ln}(M)}+{P_r}^2\ \frac{M\ T_{slot}}{{\mathrm{ln}(M)\ }{ E}}}\Bigr),
\label{Capacityequation}
\end{equation}

Where the photon energy  $E =(h.c)/ \lambda $, $h$ is Planck's constant, $c$ is the light velocity and $\lambda$ is the signal wavelength. $M$ is the modulation order. $P_r$ and $P_n$ are the received and noise power, respectively.\\
 The three terms in the denominator of the second term of  Eq.~\ref{Capacityequation}    correspond to three different capacity versus power received behavior (or regions).
The  term  $P_{n}\frac{1}{M-1}$  which   is noise-limited  results, when dominated,  in capacity change quadratic in signal power.   The quantum-limited term, on the hand,  $P_{r}\frac{1}{ln(M)}$,  leads to linear  capacity.  The   bandwidth limited  term, ${P_{r}^{2}}\frac{MT_{slot}}{ln(M)E}$ leads to  capacity saturation when dominating.

\vspace{0.5cm}

\begin{equation}
M = {2}^{k_{b}}
\label{M}
\end{equation}

\begin{equation}
T_{s} = \frac{k_{b} }{ R_{b}}
\label{Ts}
\end{equation}
  
\begin{equation}
T_{slot} = \frac{k_{b} }{M\ R_{b}}
\label{Tslot}
\end{equation}

where $k_{b}$ is the number of bits and $R_b$ is the data rate

\subsubsection{Poisson channel}
When \acrshort{ppm} is considered, the channel model from the  transmitter to the receiver is a discrete channel model, particular instance of this channel, is \acrfull{dmc}. In \acrshort{dmc}, the output of the channel depends only on the input of the channel at the same instant and not on the input before or after, it is characterized by the channel (transition) probabilities. The transition probability represents the conditional probability that the channel output is Y = yj given the channel input X = xi, it can be derived from statistical optical channel models by means of conditional \acrfull{pdfs}. The optical channel models for the DS could be poisson or \acrfull{awgn}, but for high photon-efficient DS links based
on photon-counting receiver~\cite{hemmati2011deep}, the poisson channel model is
the most adequate. its \acrshort{pdfs} is given as follows:

\begin{equation}
f_{Y|X}(k|i) = \frac{K_{i}^{k}e^{-K_{i}}}{k!} , i = 0,1. \hspace{10pt} k = 0,1,2... 
\end{equation}

where $Ki$ is the average number of photons detected when
$X = i$. If the sizes of input and output alphabets are the same, and transition error probabilities are equal, the corresponding \acrfull{dmc} is known as
M-ary symmetric channel. If we let the size of output
alphabet goes to infinity, the corresponding channel model is known as M-ary input unconstrained output channel. The decision rule that minimizes average symbol error probability, is when all input symbols are equally likely, which is known as the maximum-likelihood (ML) decision rule. In the absence of background radiation, the symbol error probability expression for \acrshort{ppm} is: 
\begin{equation}
 P_s = (M-1)e^{-K_{s}}/M   
\end{equation}

\begin{figure}[H]
\centering
\includegraphics[width=0.8\linewidth]{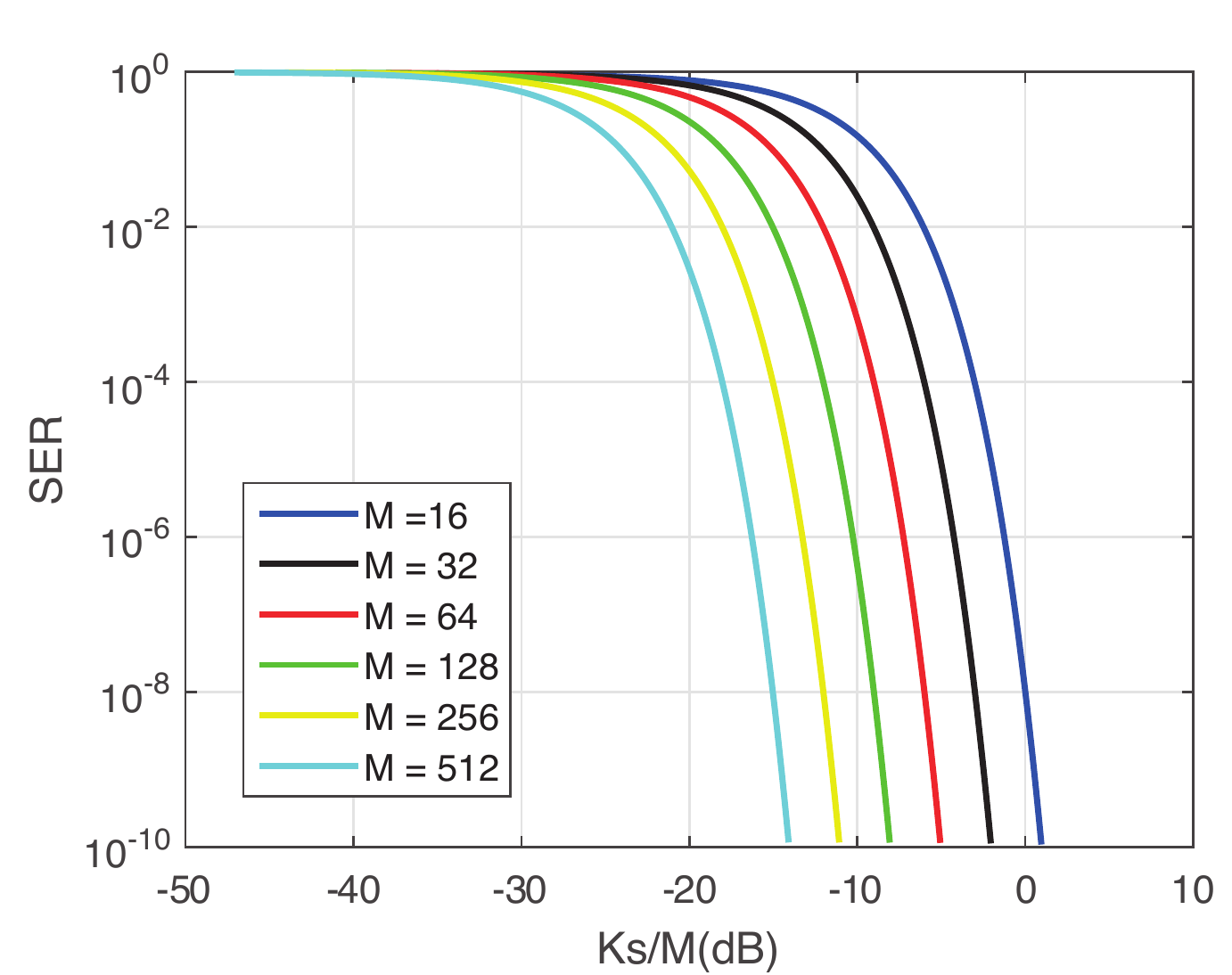}
\caption{ BER of PPM for different $M$ order.}
\label{BER}
\end{figure}

$Ks = Ml_{s}\tau_{s}$, $M$ is the \acrshort{ppm} order, $l_{s}$ is the photon flux per second and $\tau_{s}$ is time slot. As seen in Fig.~(\ref{BER}), The power efficiency can be improved by increasing the number of slots $M$, and it seems that operating in the range of $Ks/M = -15$ to $0 dB$ gives better \acrfull{ber}, where the ratio $Ks/M$ is actually the energy detected by the receiver~\cite{hemmati2011deep}.

\subsection{Coding }
 \label{sec:dsPMod2}
 
\begin{figure}[t!]
\centering
\includegraphics[width=0.7\linewidth]{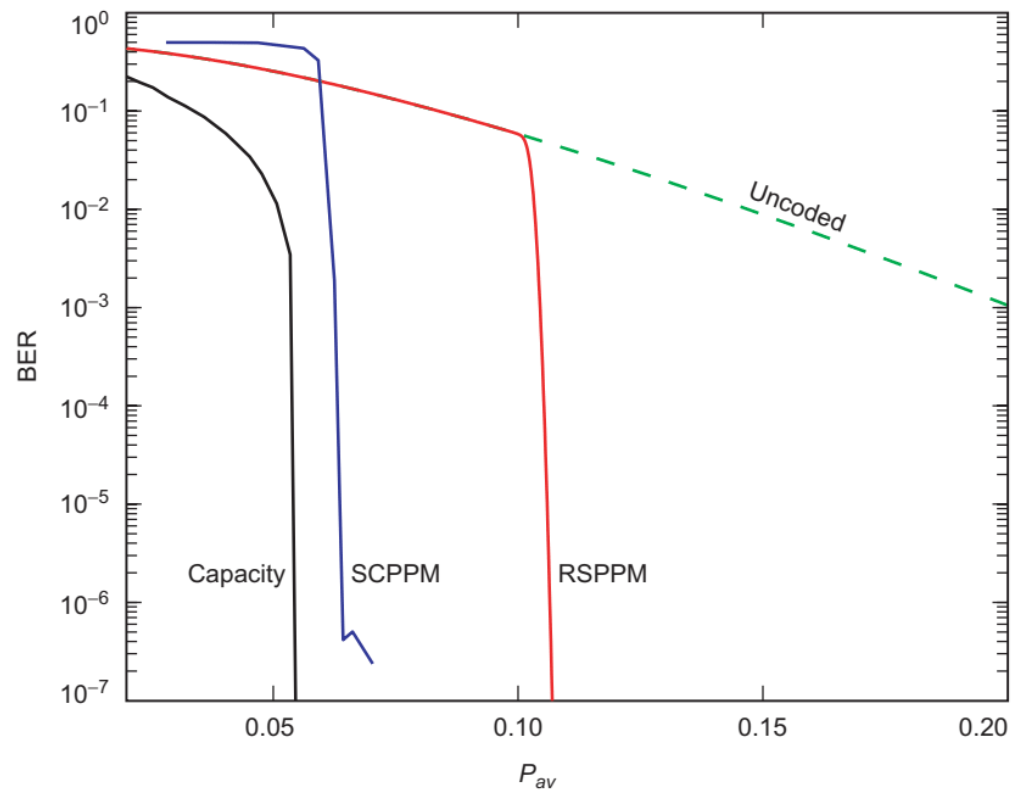}
\caption{Comparison in terms of error rate between coding techniques compared to uncoded communication and capacity limit~\cite[Fig.4-35]{hemmati2011deep}.}
\label{Fig.coding}
\end{figure}

Channel coding is a very important block in the transmission chain to ensure the reliability of \acrshort{ds} communication. 
Typically, applying ideal \acrfull{ecc} can lead to reaching the capacity limit, i.e, the maximum possible achievable capacity. However, such codes are hard to construct. Therefore, it is important to take into account the code efficiency when designing the communication links~\cite{Moision2012Designtool}. 
The code efficiency refers to the difference between the capacity limit and the error code threshold at a specific \acrfull{ber}. The smaller  the code efficiency, the closer the link capacity to the capacity limit  is.
As a reference, an uncoded system would have a code efficiency in the order of $5~dB$~\cite{moision2005coded}. 
In the early missions to space in the sixties, a decision was needed on whether to rely on block codes or convolutional codes~\cite{massey1992deep}. For marinar '69, block codes where the best options at that time given sequential decoding implementation was not that efficient at that time. Such choice led to a poor data rate. Thanks to the efficient implementation of sequential decoding, convolutional decoding has been adopted starting from Pioneer 9. Such a change offered a better coding gain of $3.0~dB$ compared a $2.2~dB$. Since then convolutional codes under different variants has been adopted for \acrshort{ds} communication such as Turbo code~\cite{divsalar1995multiple}. 
Reed-Solomon codes are commonly used with convolutional codes for DS application as it is efficient in correcting complex and bursty errors~\cite{de2011reliability,hemmati2006deep,hemmati2011deep,mceliece1994reed}, and has a code efficiency in the order of $2.5$ to $3~dB$ when its being used with \acrshort{ppm}~\cite{hemmati2020near}.  
Nevertheless, the most recommended and efficient coding that can be used in DS systems is the \acrfull{scppm}~\cite{cheng2006sat05,SCPPM4DS2006,moision2005coded}. This coding techniques has an efficiency between $0.5$ and $1~dB$~\cite{hemmati2020near,moision2005coded}. In Fig.~(\ref{Fig.coding}), we show the bit error rate of SCPPM compared to the capacity limit, Reed-Solomon with PPM and uncoded \acrshort{ppm}. 

Recently, there has also been some suggested codes in the field. For example the authors in~\cite{divsalar2020wavelength} show that coded optical ~\acrfull{wdma} with \acrshort{ppm} can be a good candidate for DS. 
Using the coding gain metric, which is the power gain over uncoded case, WDMA has a $7~dB$ gain for a $0.8$ coding rate, and a $12~dB$ coding gain for a coding rate of $1/6$ at a frame error rate  (FER) of $10^{-6}$~\cite{divsalar2020wavelength}.
Since fixed-rate codes like \acrshort{ldpc} and Turbo codes are not efficient for DS communication since it fails to adapt with the changes in \acrshort{ds} communication link due to time variability, authors in~\cite{liang2020raptor} propose the use of rate-less codes (e.g. Raptor codes), and hence achieves a better performance than LDPC and Turbo codes. The authors propose to further combine between spinal and polar coding to achieve even better performance.

The reliability of DS link is explored by a number of other techniques in addition to the coding. We next address the  important aspects of  \acrshort{ds}  networking protocols  and some related challenges. 


\subsection{Detection}
\label{sec:dscomRec}

Advances in detection technology play a role in providing efficient \acrshort{ds} communication. A detection method can either be coherent or non-coherent. Coherent  receivers, for example heterodyen and homodyne, have a higher spectral and dimensional efficiency, which is the capacity to use different modes (degree of freedom). 
However, they suffer from limited photon efficiency, but tend to be more practical when very high data rates are required.  
Non-coherent receivers, on the other hand, are energy efficient, but that comes at a cost of exponentially decreasing dimensional efficiency~\cite{dolinar2011approaching}. 

Rendering to their energy efficiency, non-coherent receivers are commonly used in DS such as Direct detector~\cite{hemmati2020near,caplan2007laser}.
There are different types of direct detectors among which are \acrshort{apd}, PIN which stands for p-type, intrinsic and n-type semiconductor regions in the diode structure, and \acrshort{pc}.

Due to their high photon efficiency comparing to other conventional direct detection methods, \acrshort{pc} detectors are getting a great attention and recommended, along with PPM modulation, to be  used in the future \acrshort{dsoc}. These detectors  are able to count a single  photon incident on them.  
It can be viewed as an extension of \acrshort{apd} detectors with infinite gain in which a digital output signal is generated for each detected photon~\cite{caplan2007laser}. They can offer up to $20 dB$ gain over APD~\cite{hemmati2011deep}. 
However, \acrshort{pc} detectors faces many challenges. 
Typically, signals incident on \acrshort{dsoc} detectors are faint, not only because of the huge distances and associated losses, but also due to constraints on spacecraft power and laser electrical-to-optical conversion efficiency. 
Therefore, to improve communication reliability detector {sensitivity} combined with high quantum {efficiency} at wavelengths of interest, are highly sought after. Here, \acrshort{pc} efficiency is the probability that the detector generates an electron corresponding to the detected photon.
Furthermore, DS optical links are rendered viable by reducing the duty cycle of the laser transmitter so that the power peak of the laser pulse can overcome the huge space loss, as well as associated system and channel losses, so that a few signal photons arrive at the detector. This calls for temporally narrow pulses, typically  in the order of few $ns$ long ~\cite{lyras2019deep}.\\  


The noise sources that affect \acrshort{pc} detection process are the $i)$  background noise, which comes from diffused energy from the sky, planets, or stars in the field of view of the receiver, $ii)$ dark counts electrons generated by the photon detectors, and $iii)$ thermal noise of the receiver~\cite{lyras2019deep}. In addition to the noise, receivers implementing PC detection schemes suffer from the following two losses
\begin{itemize}
    \item {\em Blocking loss.} Due to the binary nature of the detection output, noise in the detection process appears in the form of dark counts or varying detection efficiency. As a result, the detector gets blind after a detection event missing all incident photons until reset, leading to a blocking loss. Hence, the detector is limited to counting at most one photon per reset time or dead time~\cite{moision2003downlink}.

\item {\em Jitter loss.} The random delay from the incident of a photon on a detector to the time an electrical output pulse is generated in response to that photon~\cite{moision2008Farrjitter} is called a jitter loss.
A significant loss occurs if the standard deviation of the jitter is in the order of the slot width of the pulse.
\end{itemize}

\acrshort{ppm}-\acrshort{pc} technologies have been demonstrated by NASA to achieve data rates of several hundred megabit per second, e.g. $622~{Mbps}$ with a $3.8$ \acrshort{ppb} sensitivity in the Lunar Laser Communication Demonstration  and $781~{Mbps}$ with a sensitivity of $0.5$ detected \acrshort{ppb}~\cite{kakarla2020one}.

We provide next examples of \acrshort{pc} detectors that can be used for \acrshort{ds} applications. 
\begin{enumerate}
    \item The intensified photo diode that uses wavelengths of $1064$ and $1550 nm$ with minimum slot widths of approximately $500 ps$, and supports $200 Mbps$  data rate.
    \item Niobium nitride,  superconducting nanowire  customized  arrays~\cite{stern2007fabrication} support slot widths of down to $200 ps$ with detection efficiencies of $50-78\%$~\cite{rosfjord2006nanowire}. To achieve detection efficiency above $50\%$   detector arrays need to be implemented~\cite{hemmati2011deep}. 
    \item  {Superconducting nanowire single photon detectors~\cite{JPL2021SuperNano}.} 
\end{enumerate}

\section{Space protocols}
\label{sec:dscomSP}

In addition to the different physical layers techniques such as modulation, coding and detection, network protocols are of paramount importance to make the communication links reliable and overcome the different complex challenges. 
Hence, seamless connection and coordination between satellites, ground communication systems on Earth and \acrshort{ds} probes at the final terminals are needed. 
In such complex and challenging environment, due to losses, extreme distances and space characteristics, matured terrestrial networks protocols are of no use, hence calling for well tailored network protocols, called \acrshort{ds} internetworking  \cite{hooke2001interplanetary}.

In this perspective, \acrshort{nasa} for example relies on \acrfull{dtn} as an intuitive solution for space missions since these networks can handle excessive delays and fast changing environment, and yet maintain reliability.   
To coordinate the networking efforts between the different space agencies, \acrfull{ccsds} has been working on different protocols and architectures such as \acrshort{scps}
architecture~\cite{akyildiz2004state}, space packet protocol architecture~\cite{CCSDSspacepocket}, several internet protocols, and especially the  \acrshort{dtn} architecture~\cite{burleigh2003delay,CCSDSspaceprotocals,papastergiou2014delay}.  

With early missions to space, point to point or single relay links was possible especially link to low-Earth orbit. 
However, with missions to deeper space, \acrshort{dtn} architecture, thanks to the automatic store-and-forward feature, becomes a natural choice due to its  ability  in handling long and variable delays, and link intermittent connectivity issues~\cite{jones2006routing}. Automatic store-and-forward mechanism not only allows to forward packets to next destination, but also to keep a copy for future use if packet was not successfully delivered~\cite{vasilakos2016delay}.  

Current and future internetworking protocols  have many challenges to overcome  such as network topology dynamicity, links and entity heterogeneity, \acrshort{dtn} scalability and also the ability to support possible future services. Some of these concerns have already been raised such as the need for future services such as broadcast, multicast and streaming~\cite{zhao2018network}.

\begin{figure}[t!]
    \centering
    \includegraphics[scale=0.4]{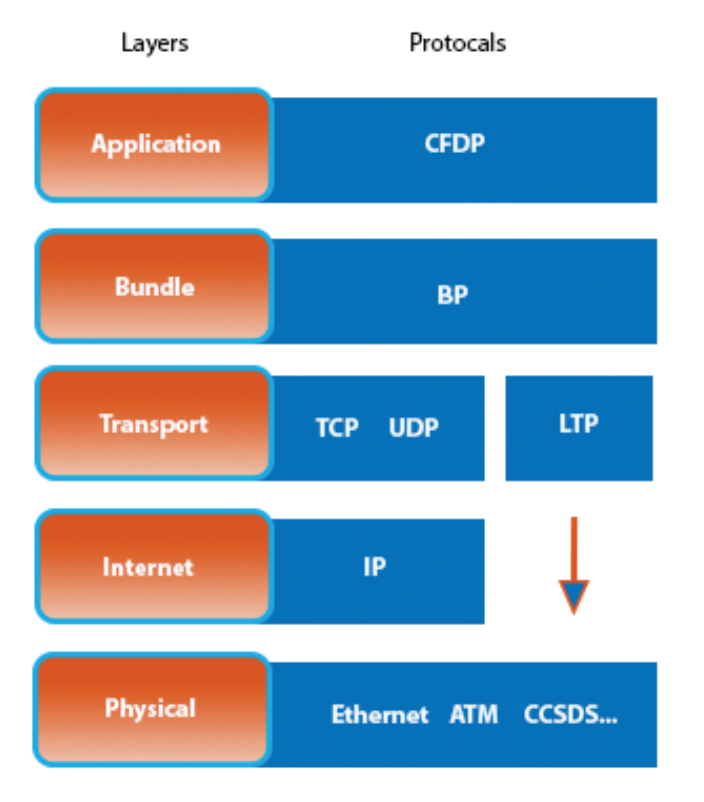}
    \caption{DTN Architecture~\cite{zhao2018network}.}
    \label{DTNArchpsd}
\end{figure}

\subsection{Discussions and outlook}
In this section, we discussed the potential modulation and coding schemes for \acrshort{ds} communication. It is worth reiterating that yet experimentation of the efficiency of existing schemes needs to be conducted for \acrshort{dsoc}. The selection of the \acrshort{ppm} order needs to be thoroughly studied and carried based not solely on the achieved capacity but rather on other criteria such as \acrshort{ber}, transmitter and receiver complexities, link gains, etc. Similar studies are needed from coding perspective to explore the possibility of new codes developed for DS communication. 
DS systems are expected to operate at speed of  several tens of gigabits per second and beyond and hence will require a major improvement over existing receiver technology both in terms of data rate and sensitivity.
 Even though coherent receivers can in principal provide near-quantum-limited receiver sensitivity, such a performance is achieved by strict conditions, which make direct detection receivers simpler to achieve and in general provide better performance~\cite{hemmati2006deep}.
As a result, in addition to  their energy efficiency and simplicity, non-coherent detectors are preferred over their counterpart. Nevertheless, coherent detection has recently started to get more interest   due to their ability to increase the capacity in \acrshort{ds} communication~\cite{kakarla2020one}.
Based on~\cite{stevens2008optical}, while using a single-quadrature  homodyne  receiver (without a pre-amplifier) resulted in a sensitivity of $1.5 PPB$ at $156 Mbps$, demonstration       with
erbium-doped fibre amplifier  coherent receivers that is pre-amplified results in $10Gbps$ with only $0.51$ \acrshort{ppb} increase in sensitivity ~\cite{geisler2013demonstration,lavery2012realizing}.  
At the network protocol level,however,  there is a need for advanced \acrshort{dtn} protocols tailored for DS links accounting for the high delays and intermittent losses.

\section{Capacity limit}

In order to decide on the modulation, coding, and detection strategies,  channel capacity needs to be quantified. 
The Shannon capacity gives the  the maximum rate in a communication channel in classical information theory~\cite{shannon1949Noise}.
In the context of \acrshort{dsoc}, the capacity is characterized by Holevo information which is based on quantum mechanic theory. The analytical closed form expression of the quantum capacity is derived based on Holevo theorem~\cite{dolinar2011approaching}, which is, asymptotically,  formulated as ${c_{d}}^{Hol}=ec_{p}2^{-c_{p}}$ where $e$ is exponential constant, $c_{d}$ is the dimensional efficiency expressed as mode/bit and $c_{p}$ is information efficiency in photon/bit unit.
In theory  optimal
coherent-state modulation achieves the ultimate quantum limit but practical sub-optimal coherent receivers are operating in near limit but in the low photon efficiency region as they hit a birck wall of ~1.44 bits/photon and by that are not efficient for \acrshort{ds} application~\cite{dolinar2011approaching}.

\begin{figure}[htbp]
\centering
\includegraphics[width=0.9\linewidth]{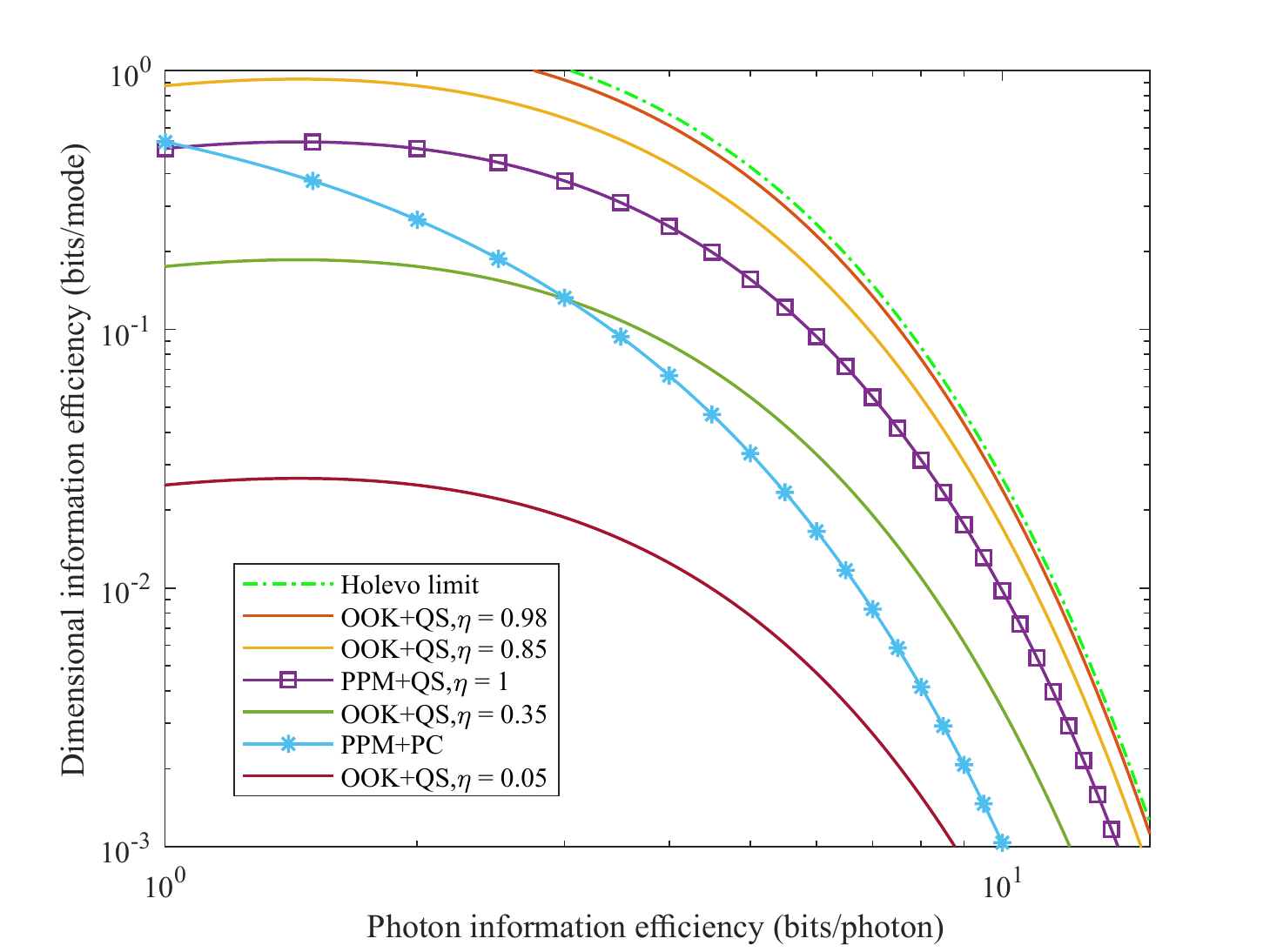}
\caption{Capacity limit}
\label{HalevovsQSvsPPM}
\end{figure}

In the case of \acrshort{ppm}-\acrshort{pc} receivers, the  capacity can be approximated with respect to Holevo limit as  follows~\cite{dolinar2011approaching,banaszek2020quantum}.

\begin{equation}
    C_{Holevo} = 2.561 C_{PCR}
\end{equation} 
proposes using coherent states with \acrshort{pc}  to get closer to Holevo limt as its easy to produce coherent states but it challenging and very impractical. Instead, suggests using quantum \acrfull{ns} communication, where number state is a  quantum state well-defined number  of photons, with the fact that quantum states are hard to produce, \acrshort{ns} is more practical then the former method and could  reach to the Holevo limit~\cite{dolinar2012fundamentals}. Nonetheless, achieving so requires very high channel transmittivity which means the
probability of transmitted \acrshort{ns} photon is received at detector.
 We can see from the following formula how \acrshort{ns} communication effected by transmittivity of the channel which is ${c_{d}}^{NS} = {c_{d}}^{Hol}(F)$, where  $F$ is multiplicative factor, in the case of OOK $F = 2^{f(\eta)}$ where ${\eta}<<1$ $f(\eta) = \eta/e$ and when $\eta{\sim}1$, $f(\eta) = ({\eta}-1)^{(\eta-1)}/e^{1-{\eta}}$. but for PPM $F = \eta/e$. As shown in Fig.~(\ref{HalevovsQSvsPPM})  hence, \acrshort{ppm} will not reach the holevo limit, even when $\eta = 1$.
 With a transmittivity of one,  NS along with \acrshort{ook} gets to the ultimate limit.
 In Fig.~(\ref{HalevovsQSvsPPM}), in addition to Holevo limit,  we show the capacity limit for \acrshort{ppm} plus \acrshort{pc} and quantum number state  modulation with OOK and PPM. 
The figure demonstrates how as quantum number state  transmittivity ($\eta$) gets higher, the capacity gets closer to the Holevo limit in the case of \acrshort{ook}.

\begin{table}[htbp]
\caption{Capacity for different planet's distances}
\begin{center}
 \begin{tabular}{||c c c c c||} 
 \hline
 Average & Planet & Received &  Reached & Average \\ 
Distance &        &   power  &  capacity & delay \\ 
   (Km)    &       & ($\mu$W)&  (Mbps)& (Minutes)\\[0.5ex]
 \hline\hline
 58 million & Mercury & 5.1856e-6 & 139.45 & 3.22 \\ 
 \hline
 225 million & Mars & 4.5053e-6 &  123.42 & 12.5 \\
 \hline
 778 million & Jupiter & 1.7741e-6 & 52.527 & 43.22 \\
 \hline
 1.2 billion & Saturn & 9.2772e-7 &  28.173 & 66.66 \\
 \hline
 4.5 billion & Neptune & 7.8951e-8 &  2.4598 & 250(4h) \\ 
 \hline
 5.9 billion & Pluto &   4.6219e-8 & 1.4409 & 327.77(5.46h) \\ [1ex] 
 \hline
\end{tabular}
\label{capacity}
\end{center}
\end{table}

In Table~\ref{capacity}, we show the achieved   soft capacity which is based on the probability as a result of counting incoming incident photon that indicate if there is a pulse or not~\cite{dolinar2012fundamentals}. For one PPM-PC detector at the average distance of various planets. We assume the transmitter and receiver diameters to be $0.22m$ and $4m$ respectively, $4W$ transmitted power, $0.25ns$ time slot, modulation order 16, and noise power reaches to $1.1620e-16$. And path loss for example for the average distance of Mars is $~366dB$ , the sum of other loss for the received power not including the detection loss is $~6.6dB$, and the detection loss is $4.35dB$.

Noticing how the capacity dramatically gets smaller as we get to a further planet distances. Things gets even worse as the planets get further far from the Earth.


 \renewcommand{\figurename}{Fig.} 
 \captionsetup[figure]{labelformat=parens, labelsep=colon}
 \chapter{DS Communication Advances}
\label{ch:dsAdvances}

\section{Orbital Angular Momentum}\label{OAM}

To better utilize the  tremendous capacity offered by \acrshort{fso}, a great deal of attention is being giving for exploring the potentials of merging the~\acrshort{fso} with  \acrshort{oam}  which emerges  as a technology to  solve the high date rate  demand of future  near-Earth and DS communications. 
Thanks to the fact that the helical wavefront of a photon is distributed in infinite ways (or modes)  spatially, allowing waves with various phase fronts to be multiplexed and transmitted over the same frequency spectrum since they can be distinguished after they propagate in space. Next, we  show the mathematical representation of \acrshort{oam} modes.

\begin{figure}[htbp]
\centering
\includegraphics[width=0.7\linewidth]{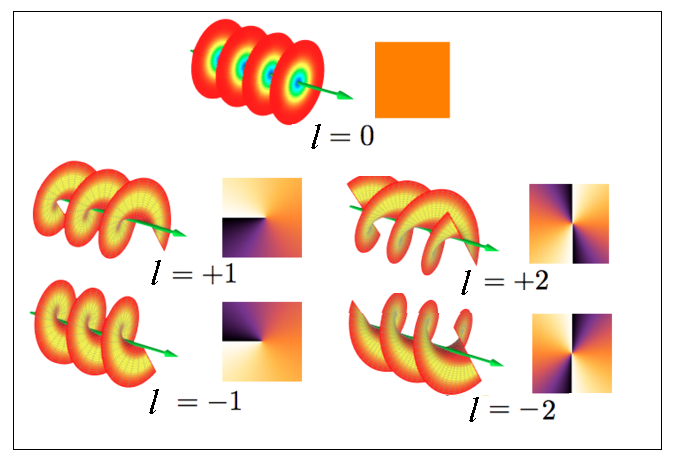}
\caption{OAM phase structure. Source: Adapted from~\cite{Wikipedia}}
\label{OAMmodes}
\end{figure}

The mathematical description of  the electromagnetic wave in the cylindrical coordinate is expressed as the \acrfull{lgb} given  by~\cite{espinoza2015optical} as:

\begin{equation}
\begin{split}
E(r,\phi ,z){\ =\frac{C^{\lvert{l}\rvert}}{w(z)}}\ {\Bigl(\frac{\sqrt{2}r}{w(z)}\Bigr)}^{\lvert{l}\rvert}{\exp \Bigl(\frac{{-r}^2}{{w(z)}^2}\Bigr)\ }{Lp}^{\lvert{l}\rvert}\Bigl(\frac{{2r}^2}{{w(z)}^2}\Bigr)\\
\times {\exp \Bigl(-i\frac{kr^2z}{2(z^2+{z_R}^2)}+il\phi +i(2p+\lvert l \rvert+1) {\arctan \Bigl(\frac{z}{z_R}\Bigr)\ }\Bigr)\ },
\label{eg:3}
\end{split}
\end{equation}

Where r, $\phi$ and z are cylindrical coordinates, $\phi$ is the azimuthal phase profile taking a value from 0 to 2$\pi$. $l$ is the azimuthal index specifies the mode which translates to the number of twists per wavelength. The sign of the mode (positive or negative) represents the direction of twist (clock or counter clockwise respectively), $p$ is the radial mode index which indicates the polarization of the beam. $ z_{R} = \pi w^{2}_{0}/\lambda$, where $w_0$ is the beam waist at $z = 0$, is the Rayleigh range which is the distance where the cross section of the beam is doubled. $w(z)= w_0\sqrt{1 + (z^{2})/(z^{2}_{R})}$ is the radius of the beam at $z$. ${L_{p}}^{\lvert{l}\rvert}$ is the associated Laguerre polynomial,  $C^{\lvert{l}\rvert}$ is a normalization constant, and $\arctan(z/z_R)$ is the Guoy phase shift which is $n\pi/2$ axial shift that a converging light wave experiences as it passes through its focus, where  n is a dimension that equals 1 in case of cylindrical wave and equals 2 when the wave is
spherical~\cite{feng2001physical}.

The generation of \acrshort{oam} beams could be done using \acrfull{spps}
which is a crystal plate that has a helical shape structure converts Gaussian beams to \acrshort{oam}. 
Moreover, diffractive phase holograms, metamaterials, cylindrical lens pairs, fiber gratings, and couplers  each of which  can generate \acrshort{oam} beams. 
The q-plates  also can  convert the \acrfull{sam}  of light  to \acrshort{oam}~\cite{thakur2013twisted}. For more about generation methods see~\cite{chen2019orbital,zhang2020review}.

Detection of \acrshort{oam} modes on the other hand can be achieved using either the inverse helical phase hologram, q-plates, \acrshort{spps}, photonic integrated circuit, free-space interferometers, which an instrument for measuring optical properties, or mode sorter which is capable of demultiplexing multiple of orthogonal spatial components occupying the same space~\cite{willner2015optical}.

Specific arrangements in the generation side results into different beam types. such as  Mathieu, Bessel, Ince-Gaussian, vortex, hypergeometric, and   \acrfull{lgb}s each of which has  different characteristics.  
 Perfect vortex beams could perform better then  \acrfull{lgb}s under turbulence based on quantum communication system~\cite{djordjevic2012quantum}. 
%
 Bessel beams, on the other hand, are diffraction free and have superior {self-healing capabilities} (size preserving for example) hence have resiliency against \acrshort{at} but only for very limited distances~\cite{vetter2019realization,birch2015long,li2017adaptive}. 
 Comparing  to  Bessel beams, however,  \acrfull{lgb}s can, due to their self-healing property,   propagate farther distances and hence more roubst~\cite{mendoza2015laguerre}. 
  \acrfull{lgb}s are  the most applied type in  \acrshort{dsoc}~\cite{mendoza2015laguerre}.


\subsection{Challenges}  
 Many recent research and experimental works have, unprecedented, exhibited very significant results in the possibility of establishing   several terabit per second capacity for \acrshort{oam}-based communications~\cite{trichili2019communicating,wang2015ultra}. \acrshort{oam} does not only tremendously enhance the 
 spectral efficiency, it also  offers a great deal of  improvement for two other very important aspects of communications, namely energy and security~\cite{sun2016physical,djordjevic2017multidimensional}.

Nevertheless, there {are some}  major challenges, listed below, facing  \acrshort{dsoc} \acrshort{oam} implementation.  
For the best of our knowledge,  thoroughly discussions of  \acrshort{oam} \acrshort{dsoc} challenges and some proposed mitigation are being presented for the first time. 

\begin{figure}[H]
\centering
\begin{subfigure}[b]{0.75\textwidth}
\includegraphics[width=\textwidth]{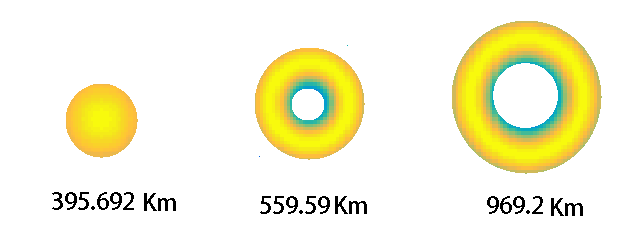}
\caption{OAM dimensions in the furthest Earth to Mars distance of 401Mkm.}
\label{AOAMmodes}
\end{subfigure}

\begin{subfigure}[b]{0.75\textwidth}
\includegraphics[width=\textwidth]{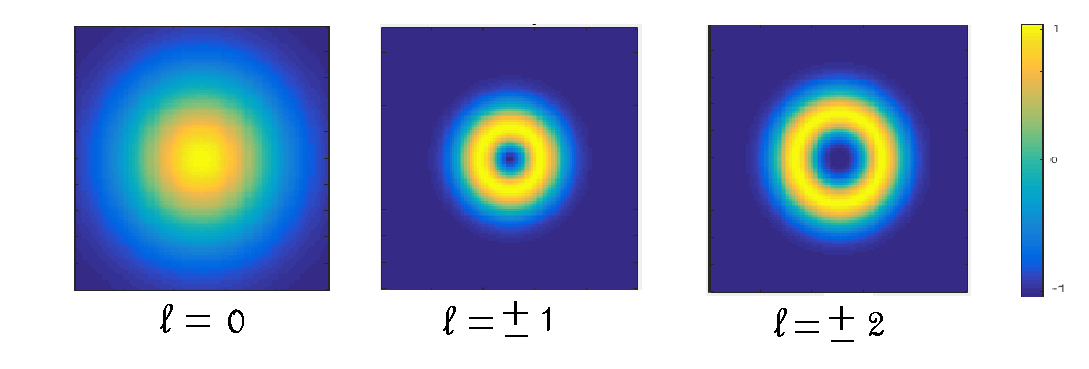}
\caption{OAM beam front distribution in the furthest Earth to Mars distance of 401Mkm.}
\label{BOAMmodes}
\end{subfigure}
\caption{OAM modes dimensions and distributions in the furthest Mars distance away from Earth }
\label{OAMmodes1}
\end{figure}

\begin{enumerate}[noitemsep]

\item Atmospheric turbulence.  \acrshort{oam} modes are very vulnerable to \acrshort{at}, as it leads to crosstalk which render the orthogonality between \acrshort{oam} states to be no longer preserved. 
Even though,   \acrshort{at} is not of a concern in  space-to-space link~\cite{kaushal2016optical}, it is the main challenge  of \acrshort{oam}-\acrshort{dsoc}    probe-to-Earth links. 

\item Spatial spread. The \acrshort{lgb}  intensity profile exhibits a donut style that gets bigger, i.e. spreads more spatially or diverge, as the beam singularity increases.   
The good news is that it has be proven experimentally that \acrshort{oam}  signals or modes, even   though diverge, preserve their orthogonality and hence can be detected even if the \acrshort{lgb} is only partially received~\cite{xie2013analysis}.  More specifically, even if a quarter or multiple of a quarter of the donut-profiled beam is captured by the receiver, \acrshort{oam} signals can still be recovered.
Considering the DS distances,  we, in Fig.~(\ref{AOAMmodes}) (a) and (b), show the dimensions (in terms of profile diameter) and spatial intensity profile  of the \acrshort{lgb}. The dimensions and spatial distribution are provided at the  farthest distance
Earth-to-Mars, namely $410MKm$ for $l=0,\Ou1$ ,$\Ou2$. As we can notice, the Gaussian beam ($l=0$)   dimension is smaller than that of the OAM-LGBs. Moreover,  the beams diverge more as their modes increase. 
In Fig.~(\ref{AOAMmodes}), is the dimension of OAM beams sent from Earth to Mars with transmitter diameter of 1m, as shown it diverse more the higher the mode number selected. In Fig.~(\ref{BOAMmodes}), the \acrshort{oam} beams preserve its shape in very long distances as an indication to where to place the receiver in such long distances between the transmitter and receiver as the there is a singularity in the middle of the beam wave front, and if partial detection in this range is accomplished~\cite{padgett2015divergence}. 

\item Lateral displacement error. Keeping the \acrshort{lgb} axis and the receiving end fully aligned is difficult for an actual communication system. For long distance transmissions, this requirement is even stringent. Alignment error does not only render  optical power loss, but also,  more crucially, causes expansion of the spiral spectrum and thus crosstalk effect and mode detection difficulty~\cite{willner2016design}.
In fact, however, it has been shown that the later offset may improve  resiliency against eavesdropping~\cite{gibson2004OAMsecurity}.\\
\end{enumerate} 

\subsection{Mitigation} 
We, in Table~\ref{TAB:OAMStateofArt}, highlight OAM sate of art that address the \acrshort{oam}  achievable capacity,   detection in  presence of AT, and   partial beams reception.  Although there are significant contributions addressing the \acrshort{fso}-OAM challenges, however they are mostly limited to few kilometers distances. 
The \acrshort{oam} detection challenges  in presence of divergence issue and the possibility of recovering modes from a small fraction of the OAM beam traveled \acrshort{ds} distances, however even though imperative in the OAM technology implementation, it is still an open research problem.
Aside from~\cite{djordjevic2010ldpc,djordjevic2011deep},  there are no other references that analyzed \acrshort{oam}  challenges in \acrshort{ds}.

\begin{figure}[H]
\centering
\includegraphics[width=0.6\linewidth]{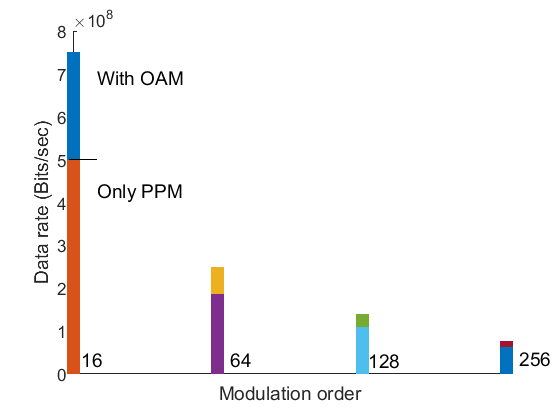}
\caption{Comparing the data rate using conventional \acrshort{ppm} vs MM modulation.}
\label{OAMPPM}
\end{figure}

\begin{center}
\begin{longtable}[htbp]{|m{3cm}| m{5.0cm}| m{1.5cm}| m{5cm}|}
\caption{OAM FSO state of art}\\    
 \hline
 References/Year & Model & Distance & Results \\ 
 \hline\hline
 \multicolumn{4}{|c|}{\textbf{Area: capacity improvement}}\\\hline
~\cite{wang2012terabit}(2012) & 16QAM polarization multiplexed with 16OAM modes and  & 1m & Capacity of 2.56 Tbit/s and spectral efficiency of 95.6bit/ s/ Hz
\\
\endfirsthead
\hline
~\cite{huang2014100}(2014) & 100 Gbit/s QPSK signal on 42 WDM channels with 12 OAM modes with polarization and wavelength  multiplexing & 1m & Capacity of  100Tbit/s\\
\hline
~\cite{yousif2019performance}(2019)& 4 OAM modes-based on MIMO/SMM making 4 OAM-multiplexed channels with each having 100-Gbit/s quadrature phase-shift keying signal (aggregate 400 Gbit/s) & 4km &Capacity of  400Gbit/s\\
\hline
\cite{wang2015ultra}(2015) & Using N-Dimentional
multiplexing and modulation link with polarization multiplexed 52 OAM modes carrying nyquist 32-QAM signals & & Capacity of 8.16 Tbit/s and ultra-high spectral efficiency of 435-bit/s/Hz \\\hline
~\cite{djordjevic2010ldpc}(2010) & LDPC-coded  OAM modulation for FSO. 
& & Capacity of 100 Gb/s could be implemented in DS \\

 \hline
 \endhead
~\cite{djordjevic2011deep}(2011) & For deep space and near-Earth communication with  LDPC-coded  OAM modulation and multiplexing as a additional degree of freedom.
& & The spectral efficiency of proposed scheme is $N^{2}log_{2}N$ times better then PPM, where N is the number of OAM modes used.  \\
 \hline 
 
  \multicolumn{4}{|c|}{\textbf{Turbulence compensation}}\\\hline
~\cite{zhou2020high}(2020)& Adaptive optics compensation approach based on the wirtinger flow (WF) algorithm.
 & 1km & The RMS error reduces to $10^{-15}$ compared with $10^{-2}$ for the conventional
gerchberg-saxton algorithm
 \\\hline
~\cite{dedo2020oam}(2020) & Using gerchberg-saxton (GS) algorithm and convolutional neural
network & 1Km to 6Km & Recognition accuracy of in high turbulence strength of 69.50$\%$ To 100$\%$ for medium and weak turbulence levels \\\hline
\cite{li2020atmospheric}(2020) & Sensorless adaptive optics (AO) and a convolutional neural network (CNN) to demodulate OAM modes & 1.2Km & BERs under strong turbulence decrease from $10^{-2}$ to $10^{-4}$\\\hline
\cite{hao2020high}(2020)& Convolutional neural network (CNN),
a deep learning (DL) technique &1Km to 6Km &Recognition accuracy  of 97.1 under moderate turbulence
and 80 under strong turbulence\\\hline
\cite{zhao2019orbital}(2019)& Diffractive deep neural
network (D2NN) 10 types of multiplexed OAM modes.&
& Recognition accuracy in high turbulence strength of 50$\%$ To 100$\%$ for medium and weak turbulence levels\\\hline
 \multicolumn{4}{|c|}{\textbf{Partial reception}}\\\hline
~\cite{xie2013analysis}(2013)& Using IR camera with SLM and a technique transform OAM states into transverse momentum states&4km& Capacity of 200Gbit/s\\\hline
~\cite{elhelaly2018reduced}(2018)& System modal for detecting a reduced area receivers &1 km&\ 12 b/s/Hz higher rate at SNR 25dB then using a complete are receivers \\
\hline
~\cite{deng2019orbital}(2019)& Using synthetic partial aperture receiving (SPAR) approach &  & Adjacent modes crosstalk could reach to 22$\%$, for non-adjacent modes crosstalk could reach as low as 7.71$\%$ \\
 \hline

\end{longtable}
\label{TAB:OAMStateofArt}
\end{center}

The authors in~\cite{djordjevic2011deep} proposed combining the \acrshort{ldpc} code with OAM-based modulation schemes to overcome \acrshort{at} effects. With their presented methods, the authors in~\cite{djordjevic2011deep}   achieved  a spectral efficiency  of $MN/\log_2(M)$ over that of \acrshort{ppm} modulation, where  $M$ is the modulation order and $N$ is number of OAM modes.\\ 
\textbf{Mode Modulation} we suggest a simple yet promising idea,which apparently turned out to be also suggested by~\cite{Yuwen2018MMOAM} which we refer to as \acrfull{mm}, to overcome crosstalk, increase power efficiency, and spectral efficiency. In \acrshort{mm}, we propose instead of multiplexing mode and having to handle the crosstalk and the degradation cause in performance and limitations it causes in transmission distances, we send only one mode at a given signaling time. 
 In addition to the information carried by an \acrshort{oam} mode, we  use the mode number to carry some information (and hence the name mode modulation). Without loss of generality, let $N=2^c$ where $c$ is integer. We can map  each   possible  $c$  bit sequence to a unique mode number and hence $c$ extra  information 
  can be carried by each mode.  For example, 
 if we use two modes  $+1$ (represented by bit zero) and $-1$ (represented by bit one) to transmit a sequence of binary bits   we only need to transmit one bit over the mode   specified by the consecutive bit. Example to send the sequence $0110$ we only send bit zero over mode $-1$ and bit 1 over mode $+1$. In this case, we have doubled the capacity without the burden of  dealing with crosstalk and cost of the power consumption. More general, \acrshort{mm} increases   spectral efficiency by $({\log_2(N)})$. To detect the signal correctly at the receiver side, however,  the mode needs to be detected as well as the data carried by the mode. In our future work we propose  to analyze the performance of the MM modulation analytically and experimentally.

\begin{figure*}[t]
\centering
\includegraphics[width=1.0\linewidth]{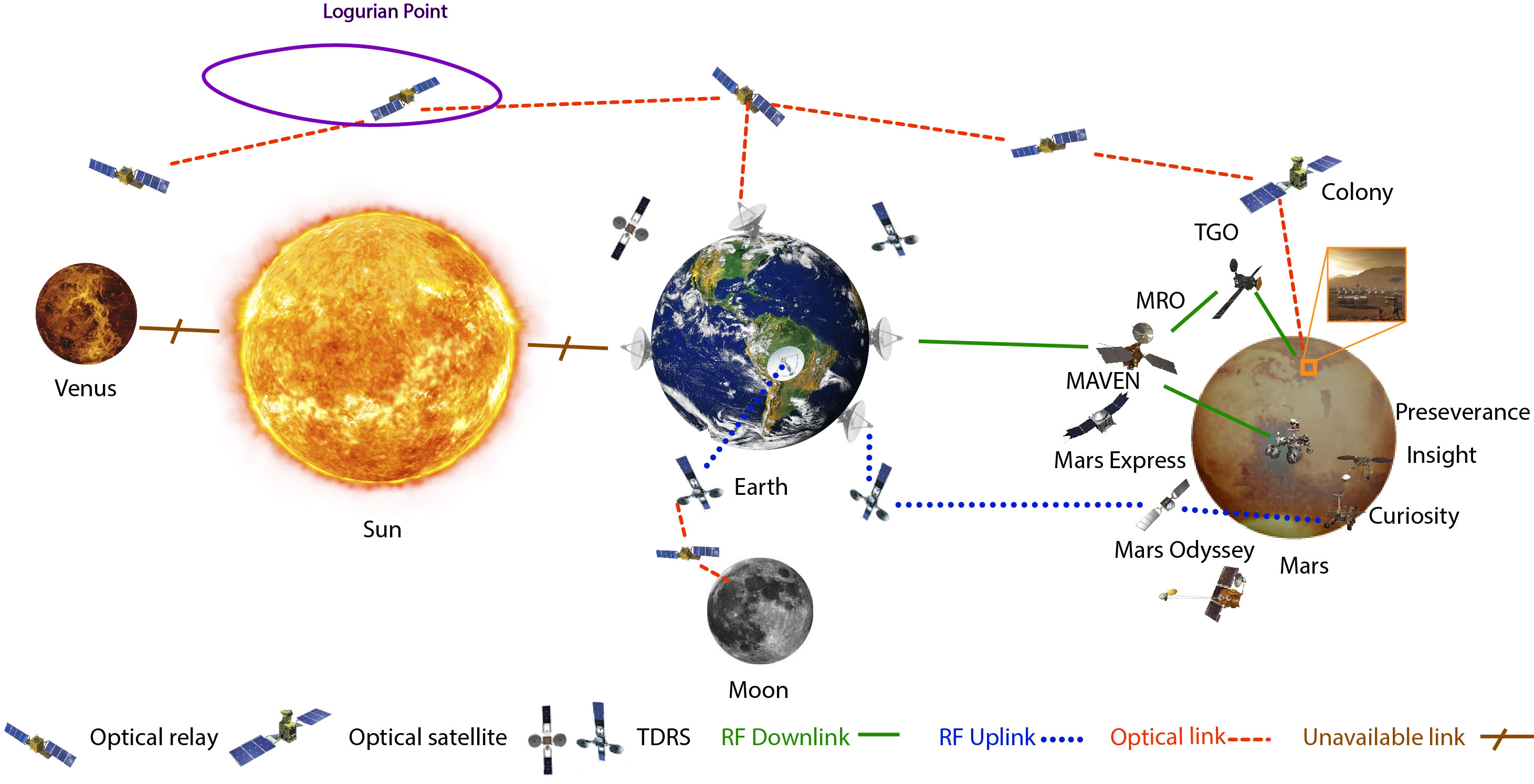}
\caption{Current  RF and future optical  interplanetary network}  
\label{ScaNnetwork}
\end{figure*}

\section{Interplanetary Network}
 \acrshort{nasa} has three operational networks that provide communication services to supported missions: DS network, near Earth network, and space network. DS network consists in multiple antennas distributed around the world. On the other hand, near-Earth-network consists of ground stations that provide communication and tracking services. The \acrfull{sn}  is a constellation of geosynchronous relays which consists of 10 satellites, called \acrfull{tdrss}~\cite{zhao2018network}. 
Connecting these  networks with the relay system distributed around space is the first step to make \acrfull{ipn}. 

 {In Fig~\ref{ScaNnetwork}, we illustrate current RF-based and future optical interplanetary network. }
In 2016, \acrshort{nasa}  lunched Next-Generation Space Relay Architecture Concept Study for 2040 to evolve NASA's space communication and navigation networks.
Planetary relays such as Mars relay is expected to operate using \acrshort{rf}  bands, mainly Ka and X bands, providing up to 125 Mbps and also optical
links that provide up to 300 Mbps. 


Potential solutions have been proposed to be implemented in the near to long term for \acrshort{ipn} with different techniques. These ideas include  optical communication terminals on planets that provide data rate  $10$ to $100$ times higher than \acrshort{rf}~\cite{alhilal2018future}. Also, other solution includes locating spacecrafts with respect to Sun-Earth's Lagrangian points to overcome obstruction or conjunction in the way of the links. However, the most interesting idea is  implementing \acrshort{ipn} with thousands of what is known as cube-sats, which are small satellites promising high speed using optical communication. These cubesats are distributed in the space promising achievable rate of 2Gbps~\cite{velazco2020inter,saeed2020cubesat}.

\section{Quantum Communications} 
\label{Quanta}
Some of the quantum properties had gain attention to be used in communication network by using the quantum states to carry information~\cite{zhang2019quantum} . 
A \acrfull{qc} is expected to break through the limit of classical communication technology in the aspects of communication security, computing power, information
transmission~\cite{zhang2019quantum}. 
However, what is unique about \acrshort{qc} is that theatrically quantum cryptography promises to obtain absolute secure channel. 
There are many techniques in \acrshort{qc}  among which are quantum key distribution \acrfull{qkd}, \acrfull{qt}, \acrfull{qss}, \acrfull{qsdc} and many more. 
 QKD is one of used protocols in free space communication as well as fiber optics that has been proven it can achieve an average secret-key rate of $47.8$ Kbits for  up to $2600$ Km distance~\cite{chen2021integrated}. 
However, a global \acrshort{qkd} and similarly sclable \acrshort{qc}  are not been realized yet mainly due to the fact that quantum repeaters ~\cite{briegel1998quantum,duan2001long} can not be  deployed~\cite{yang2016efficient}. Hopefully with ongoing research in the field, teleportation between photons without classical link restriction, distributed in the interplanetary network  would be possible and hence alleviate   the latency resulting from the long distances in communication links.

\renewcommand{\figurename}{Fig.}
 \captionsetup[figure]{labelformat=parens, labelsep=colon}
\chapter{\acrshort{dsoc} Link Engineering and Performance Analysis}
\label{lastchapter}

In this chapter we first discuss the design link parameters for PC-PM detector and conclude their values for a desired data rate, then we get a closer look on performance analysis of the link at different  characteristics.

\section{Link  Design  for PC-PPM Based on Required Performance }

Link design for DS communication in general has specific concerns.
Depending on the distance, mission requirements, capacity needed, atmospheric conditions, Sun-Earth-probe angle and Sun-probe-Earth angle,  modulation and coding used  through out the mission can be adapted to get the best performance out of the link. The capacity of the   communication link is  a function of the received and noise power, modulation type, coding scheme, and the detection method which in this analysis would be photon-counting.  \acrshort{ppm} operates efficiently at high peak to average power ratio, hence it is suited for DS environment. In this section, the method for selecting the \acrshort{ppm} order and coding to achieve  a certain data rate is described. The modal of the channel used is Poisson, which is  commonly adopted for DS optical links.  According to the quality of service (data rate in this context) that needs to be met, the required  power  is  concluded. The error correcting codes, are also chosen based on the targeted data rate. The   good candidate coding algorithms   for DS communication are the \acrshort{rs} and \acrshort{scppm}. Since the later out preforms \acrshort{rs}, as stated in chapter~\ref{chp:dscom}. \acrshort{scppm} is  chosen in this design process (similar to ~\cite{moision2003downlink,lyras2019deep}).
Depending on information provided to the decoder by the receiver,  capacity can be categorized either as hard or soft~\cite{hemmati2006deep}. In one hand, in case   the receiver estimates each \acrshort{ppm} symbol, the resulted capacity is referred to as hard capacity. This capacity is 
specified based on probability of symbol errors. 
 However, if the   receiver makes a   decision   based on counting photon per slot, 
the capacity is categorized as a soft capacity. This capacity as at least as high as as the hard as counting slot  provide more information to the decoder. 

Our analysis through out the chapter is based on soft capacity in a possion modeled channel using direct detection photo-counting detectors~\cite{caplan2007laser}. The following  equation is   Eq. (\ref{Capacityequation}), approximated soft capacity of poisson channel, which we explained  previously to give overall idea about achievable DS link capacity, as it will be used for the DS link analysis and included in producing the graphs. \\
\vspace{0.2cm}

\begin{figure}[!t]
\centering
\includegraphics[width=0.7\linewidth]{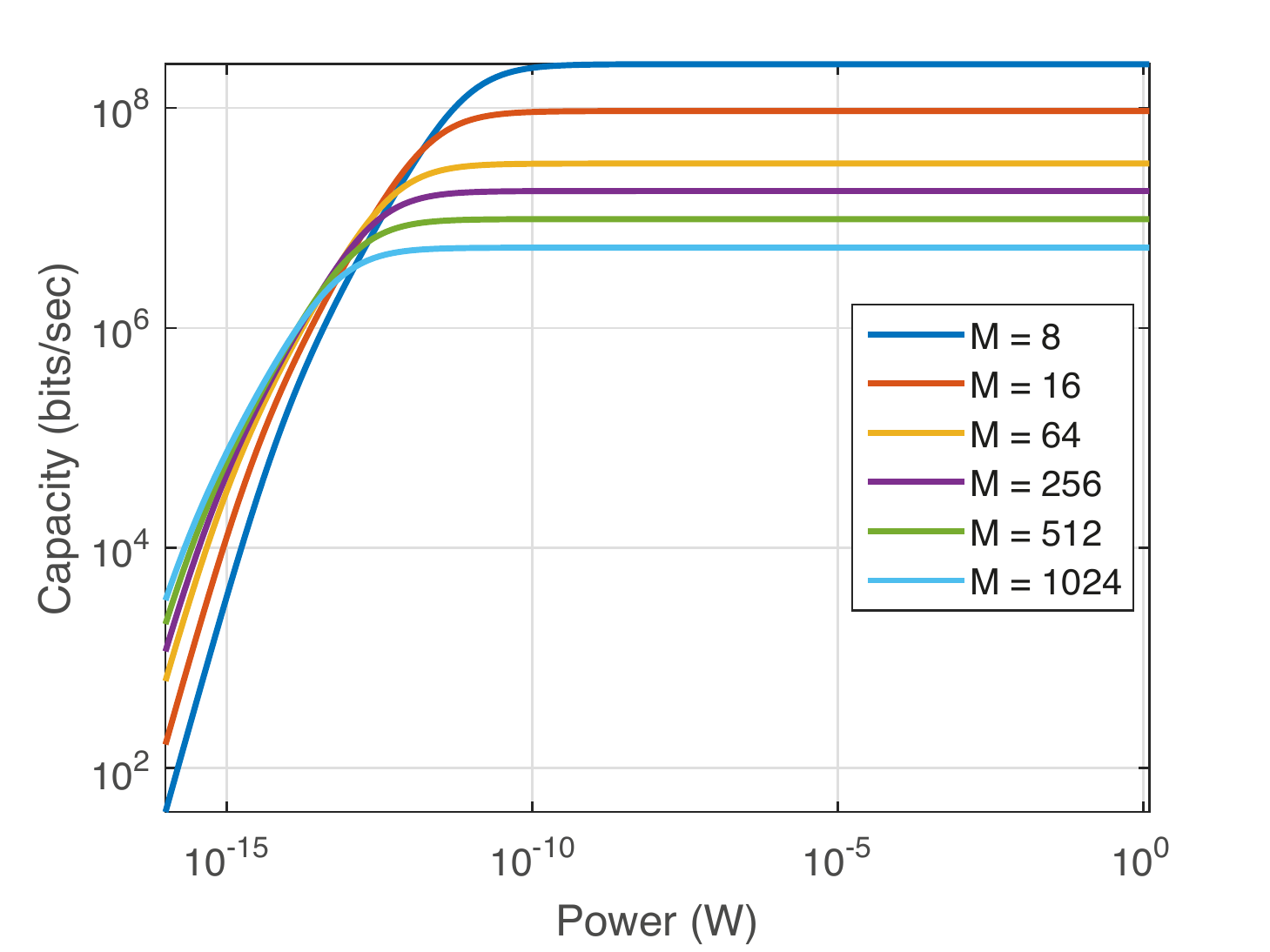}
\caption{Capacity vs received power  for different $M$ orders  }
\label{CvsPavrgCapacity}
\end{figure}

In Fig.~(\ref{CvsPavrgCapacity}) we plot  the capacity against power received for different \acrshort{ppm} order.  For low power regime,  the highest capacity is achieved by the highest $M$ order (i.e. with the modulation scheme corresponds to the slowest data rate).  As the power increases, however,   lower order schemes would be efficient to get a higher capacity Fig.~(\ref{CvsPavrgCapacity}). 

\begin{figure}[!t]
\centering
\includegraphics[width=0.7\linewidth]{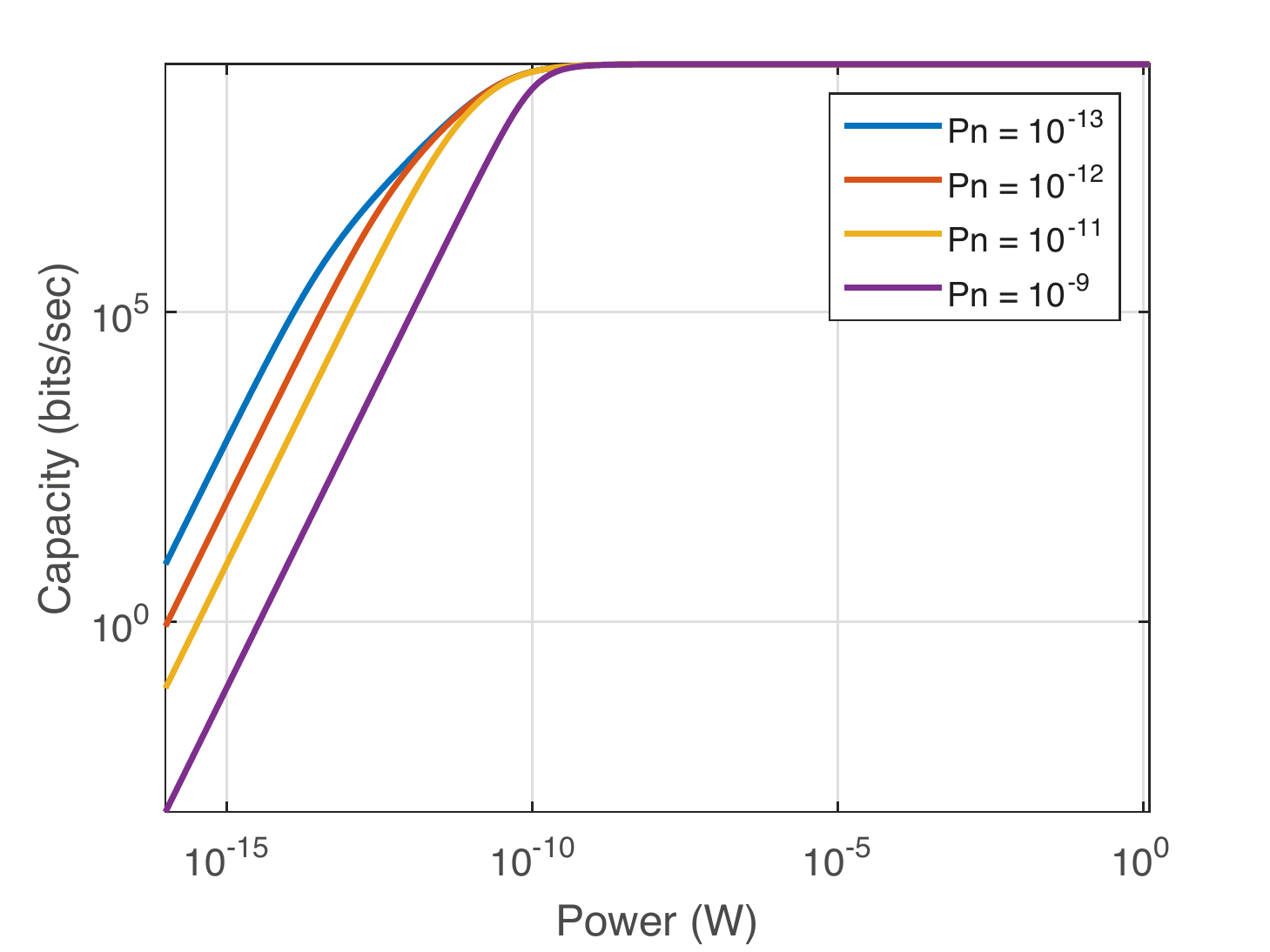}
\caption{Capacity vs received power for $M=16$ and various noise power values. }
\label{M16withnb}
\end{figure}

In Fig.~(\ref{M16withnb}), the impact of noise on the achievable capacity at different received power is shown. As depicted in this figure  in the case of $M=16$,  when power received reaches $10^{-10} W$,   noise will start not to  have an effect on   capacity.  In other words,  at some range of noise power, there is a threshold of required power that gives the maximum capacity for a given $M$ order and noise in that range will not have an effect.\\ 
It is worth mentioning that, for Mars links the $Pn$ is in the range $10^{-13}$ to $10^{-14}$ depending on the receiver used~\cite{lyras2019deep}. 

\subsection {Parameters selection}
Operation close to capacity is achieved by selecting the optimum \acrshort{ppm} order for the desired rate. 
In addition to order selection, concatenating the PPM  with an \acrshort{ecc} so that the decoded BER satisfies some threshold is crucial. The optimal order and \acrshort{ecc} rate vary with $n_{s}$ and $n_{b}$. While $n_{s}$ is the average number of single photons per
pulse, $n_{b}$ is average number of noise photons per slot time.  $n_{s}$ is  given by to $n_{s} = {l_{s}}{M}{T_{slot}}\times{\eta}$, Where $l_{s}$ is the average number of single photons incident on the detector per second, and $\eta$ is the quantum efficiency of the detector which will be discussed in the next section of performance analysis.    $n_{b}$ is given by $n_{b} = {l_{b}}{M}{T_{slot}}\times{\eta}$, where $l_{b}$ is the average number of noise photons incident per
second. 
To optimize the throughput over
the course of a mission may require an adaptive choice of a set of modulation orders and corresponding \acrshort{ecc}s \cite{hemmati2006deep,moision2003downlink}.
\subsubsection{Modulation order}
Choosing the optimum modulation  order, denoted by $M^*$,  may depend on many aspects.  
Maximizing capacity for a given received and noise power  is an important factor for design and hence the choice of $M$. The formula  below~\cite{moision2003downlink}.
\begin{equation}
    M^*\left({\mathrm{Pr},Pn\ }\right)=2^{{arg\ max}_mC(2^m,MPr,Pn)}
\end{equation}

\begin{figure}[H]
\centering
\includegraphics[width=0.7\linewidth]{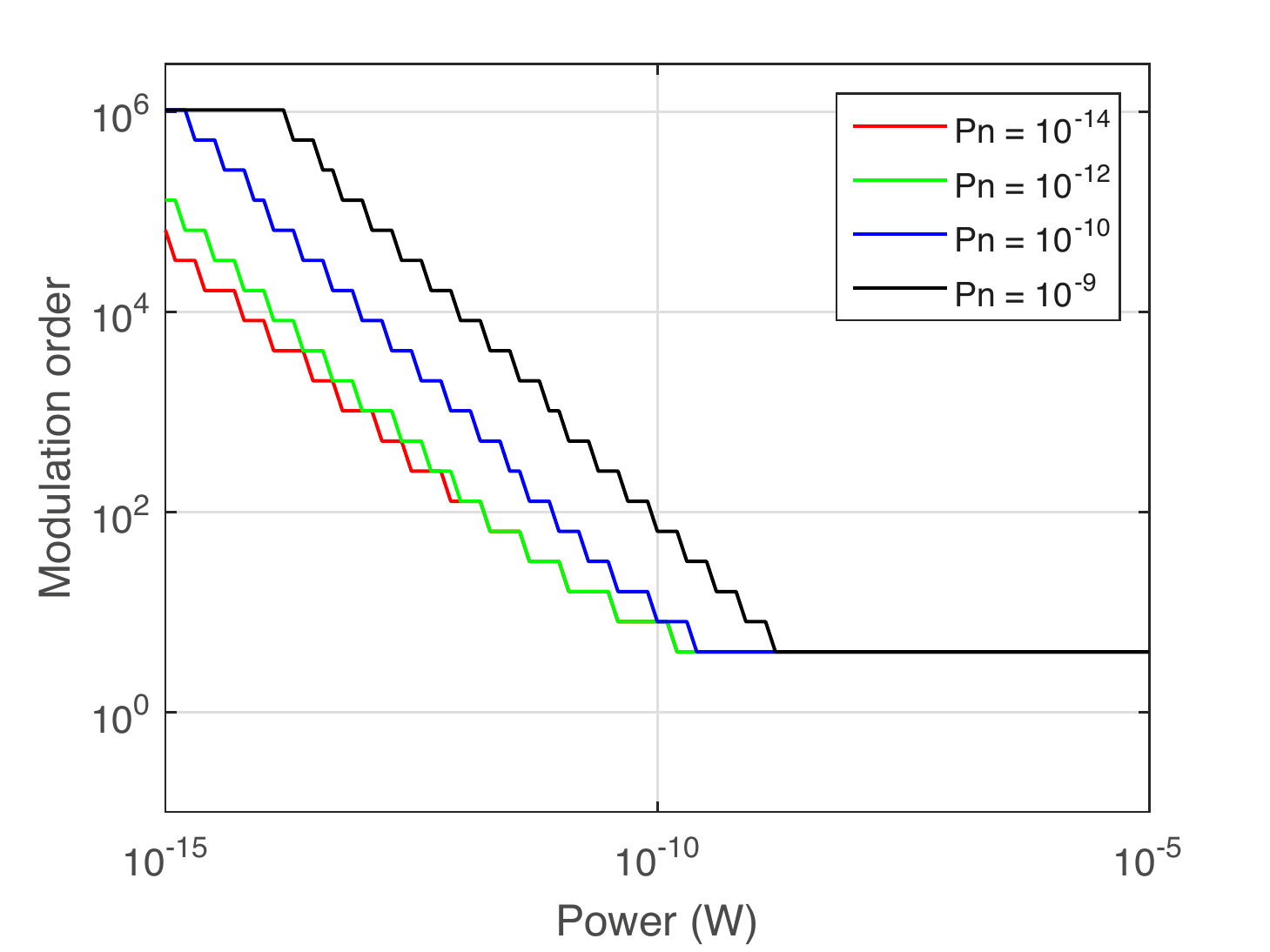}
\caption{Optimum modulation order vs received power at different noise power values. }
\label{MaiximumMoeder}
\end{figure}

Illustrated in Fig.~(\ref{MaiximumMoeder}), is optimum modulation  for different values of $P_n$.
 Switching to the maximum order that will get the desired capacity. $M$ increases with increasing $P_n$, and Similarly $M$ decreases with $P_r$.
\vspace{0.5cm} 

\subsubsection{Coding design}
 Choosing the error code ratio is a function of the desired capacity and modulation order at that capacity~\cite{moision2003downlink}.
\begin{equation}
    Rpmm\left(Pr,Pn\right)=\frac{C \times M^* }{{log}_2(M)^* }
    \label{ecc}
\end{equation}
Where the capacity unit would be in Gbps.

\subsubsection{Required power} 
In this section, we show how to make  modulation and coding design based on a certain data rate requirement and hence determine the power needed  to achieve the desired performance. 
Suppose a system with a slot width of $1 ns$ and background noise $P_{n} = 2.1\times10^{-14}$
and we are interested to find the power needed to achieve a rate of $56 Mbps$ and choose appropriate coding and modulation parameters.
From Fig.~(\ref{CvsPavrgCapacity}), we find the optimum PPM order to achieve this data rate is $M = 64$, and the minimum
required power is $10^{-12.5}$  which can be represented in single photon per slot $n_{s}/M$ which is $p_{avg}$ equals to  $0.0541$, going back to  Fig.~(\ref{Fig.coding}) in chapter~\ref{chp:dscom} . To achieve 56 Mbps, we choose a rate $R = 0.6 = 0.056/(log_{2}(M)/M)$ following Eq. (\ref{ecc}), ECC
and concatenate it with 64-PPM. The performance of two candidate ECCs for this operating point are
SCPPM($3/5$, $2$, $64$, $16410$) and RSPPM($4095$, $2457$, $64$) are illustrated in Fig.~(\ref{Fig.coding}) in chapter~\ref{chp:dscom}. Their performance may
be compared with capacity for $56 Mbps$ and the uncoded $M = 64$ performance, which, since it carries
no coding redundancy, yields $94 Mbps$. The SCPPM code operates \& $0.5 dB$ from capacity, the RS code
operates $2.5 dB$ from capacity, and  uncoded PPM $7.2 dB$ from capacity (at 56 Mbps). An appropriate comparison for uncoded $64$-PPM is with capacity for $94 Mbps$, from which uncoded performance is $4.7 dB$. (It would be more efficient to achieve $94 Mbps$ with a rate $3/5$  code mapped to $32$-PPM).

\section{Link Performance for Given Parameters}
\label{analysis}
The values used trough out this section are based on \acrfull{ccsds}~\cite{CCSDS}, for \acrfull{hpe}, with the assumption of \acrfull{celos} and the DS channel represented as a Poisson channel. Due to the  limited power nature of the link, \acrfull{im} and \acrfull{dd} with photon-counting detectors are considered for receiver architecture. Under these conditions, the \acrshort{ppm} signaling and a \acrfull{scppm} error code is used~\cite{moision2003downlink}.

Through out this section, we, unlike the previous one,  for a given  set of parameters (that may or may not be set based on \acrshort{ccsds})   tend to evaluate the DS link performance. 

\subsection{Received power – before the photon detector: }

The received power after the receiver telescope and before the photon detector Pr,ap (Watt) can be estimated as:
\begin{equation}
    {{P}}_{{r},{ap}}{=\ }{{P}}_{{t}{\ }}.{{G}}_{{t}{\ }}.{{G}}_{{r}}.{{L}}_{{fs}}.{{L}}_{{a}}.{{L}}_{{c}}.{{L}}_{{s}}.{{L}}_{{pt}{\ }}.{{\textrm{$\eta$}}}_{{t}}.{{\textrm{$\eta$}}}_{{r}}
\end{equation}

Where $P_{t}$ is the transmitted power, as defined previously, $G_{t}$ and $G_{r}$ are the transmitter and ground receiver aperture gains, respectively. $L_{fs}$ is the free space loss factor, $L_{a}$ is the atmospheric transmittance,  $L_{c}$ is the cirrus transmittance factor,$L_{s}$ is the scintillation loss factor $L_{pt}$ is the pointing loss factor, ŋt and ŋr are the transmitter and receiver efficiencies respectively. These parameters are to discussed below.

\subsubsection{Free space loss}

Free space loss due to the optical wave propagation from the transmitter to the receiver is defined as follows, where $\lambda$ is the wavelength and $R$ distance is in meters.

\begin{equation}
   {{L}}_{{fs}}{=\ }{\left(\frac{{\lambda }}{{4}{\pi }{R}}\right)}^{{2}} 
\end{equation}

\subsubsection{Transmitter and receiver gain}

The transmitter gain can be estimated using the following formula,  with $D_t$ the diameter of the transmitter aperture of the telescope:

\begin{equation}
{{G}}_{{t}}{=\ }2\times{\left(\frac{{\pi }{Dt}}{{\lambda }}\right)}^{{2}} 
\label{gain}
\end{equation}

 The equation before simplifications is $G_t = \frac{8}{{w_{0}}^2}$, where $w_{0}$ in rad is   half beamwidth angle and equals to $w_{0} = {\sqrt{8}\times{\frac{\lambda}{\pi D_t}}}$, and if the beam is Gaussian $w_{0} = \frac{2\lambda}{\pi D_t}$, which its the case in this analysis, so the gain would be Eq.~(\ref{gain}).
The majority of receiver telescopes used in optical communications have
central obstruction such as the secondary mirror of Cassegrain telescopes, to increase focal length which in return decrease the transmitted optical power and make the beam narrower. 
This type of
telescopes consists of two mirrors, a primary and a secondary. The primary mirror has larger diameter than the secondary one, while the secondary obscures the primary~\cite{klein1974optical, booth2018optical}. Receiver gain is as follows:

\begin{equation}
{{G}}_{{r}}{=\ }{\left(\frac{{\pi }{D_r}}{{\lambda }}\right)}^{{2}}(1\text{-} {{{\gamma}}_{{r}}}^{{2}}) 
\end{equation}

Where $D_r$ is the receiver’s primary aperture diameter in meters, $\gamma$r is the receiver’s obscuration ratio which is $b_r/D_r$. br is the receiver’s secondary aperture diameter.

\subsubsection{Atmospheric transmittance}

The signal is mainly attenuated due to the aerosol and molecular particles of the atmosphere that cause the
absorption and scattering of signal’s irradiance. $L_{a}$ is atmospheric transmittance factor which account the loss effects of atmosphere on the leaser beam~\cite{hemmati2006deep,alkholidi2014free,CCSDS}.

\subsubsection{Cirrus transmittance }
\acrshort{celos} transmission is assumed. However, even when the sky appears relatively clear, semi-transparent ice clouds, cirrus clouds, can be present along the path,  however, in DS analysis
cirrus clouds losses are accounted for by taking an additional power margin~\cite{degnan1993millimeter}.

\subsubsection{Scintillation loss factor}

Scintillation refers to the random optical power fluctuations caused by atmospheric turbulence~\cite{perlot2007evaluation}. In DS optical down links, due to the photon-counting nature of the link, large receiver 
apertures are targeted (more than $4$ meters) to minimize its affect~\cite{osche2002optical,giggenbach2008fading}.

\subsubsection{Pointing error loss factor}
The necessity for narrow beam width subject to the long range of a DS link makes accurate pointing critical. Pointing losses may arise due to the jitter and pointing bias. Pointing error loss can be estimated for a specific probability level, $p_0$ from the probability density function of normalized received intensity taking into account the pointing error. A pointing error of $\lambda/D$ radians, where $\lambda$ and $D$ are the signal wave length and the transmitter diameter respectively corresponds to approximately $ 85\% $  intensity reduction~\cite{hemmati2020near,klein1974optical}.

\begin{equation} 
L_{pt}= p_{0}^{{(\frac{ 4\sigma ^{2}_p}{ w^{2}_0})^{}}}  
\end{equation}

Where  $\sigma_{p}$ is the total variance of  pointing errors.
 \\

\subsubsection{Transmitter/receiver efficiencies}

The transmitter/receiver optical efficiencies are introduced as a way to  capture the losses due to the coupling of the laser beam to the optical system. Transmitter/receiver losses are denoted by $\eta_t$/$\eta_r$ ~\cite{moision2003downlink}.

\subsection{Detected power}
Detected power, $P_{det}$ for photon-counting detector are effected by blocking loss $L_{b}$ and jitter losses $L_{j}$ and also the detector’s quantum efficiency, $\eta_{det}$, which states the probability of the photon that will generate a corresponding electron.

\begin{equation}
{{P}}_{{det}}{=\ }{{P}}_{{r,ap}{\ }}.{{L}}_{{b}{\ }}.{{L}}_{{j}}.{{\eta}}_{{det}}
\end{equation}

\subsubsection{Detector blocking loss}
photon-counting detectors become inactive (blocked) for some time after the detection event, as we discussed in chapter~\ref{chp:dscom}. This blocking leads to losses  relative to an ideal detector, based on Markov model for the detector state~\cite{moision2011blocking}. Assuming high \acrfull{snr},  blocking loss$L_{b}$ for single detector which is in this case $L_{b} = \mu$ is equal to:
\begin{equation}
\mu= \frac{1}{1+l\tau}
l = l_{s}+l_{n}
\end{equation}

\begin{figure}
    \centering
    \begin{subfigure}[t]{0.46\textwidth}
     \centering
    \includegraphics[width=\textwidth]{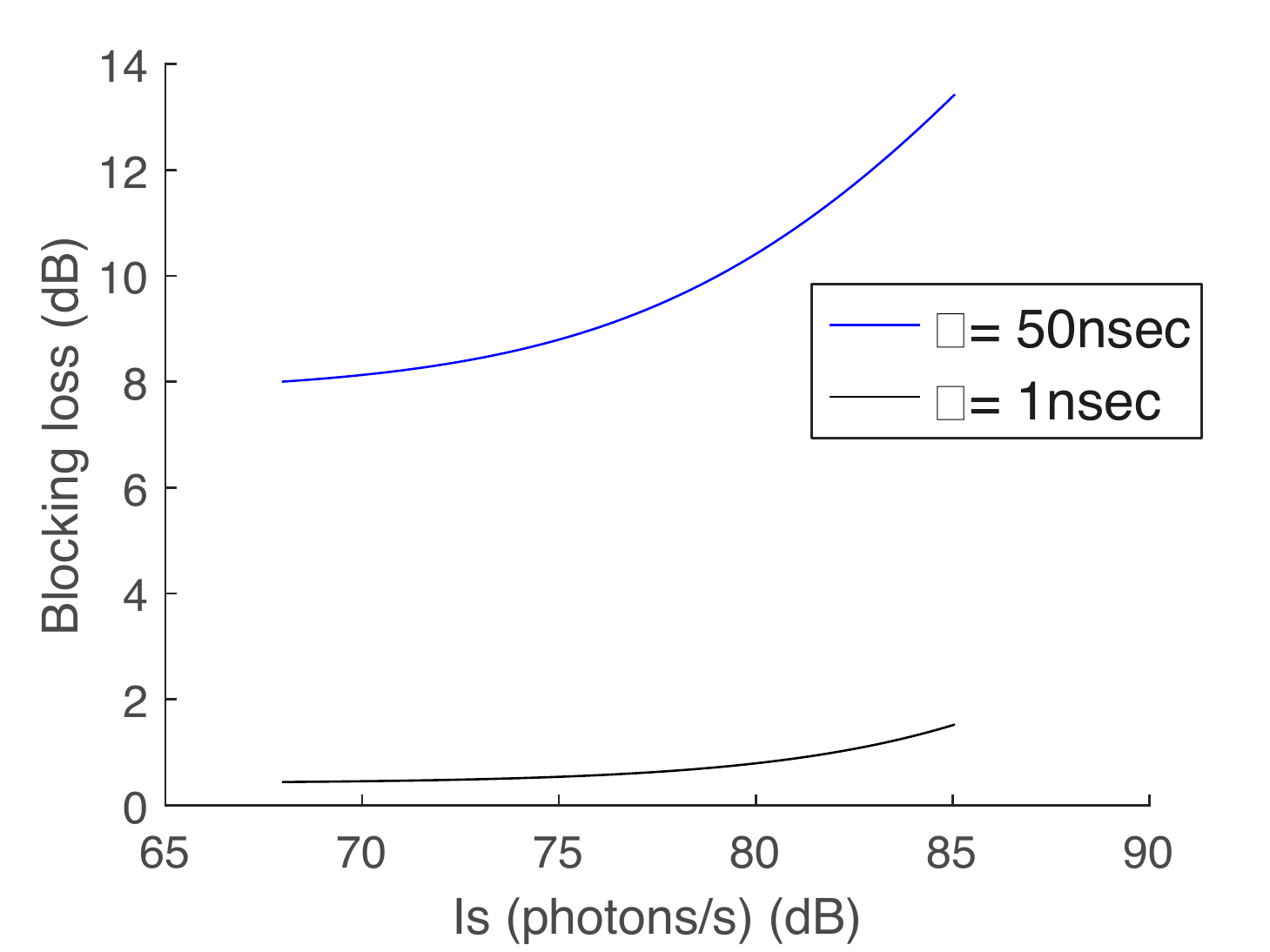}
    \caption{Blocking loss for different power received values}
   \label{blossnormal}
    \end{subfigure}
  \hfill
      \begin{subfigure}[t]{0.46\textwidth}
       \centering
    \includegraphics[width=\textwidth]{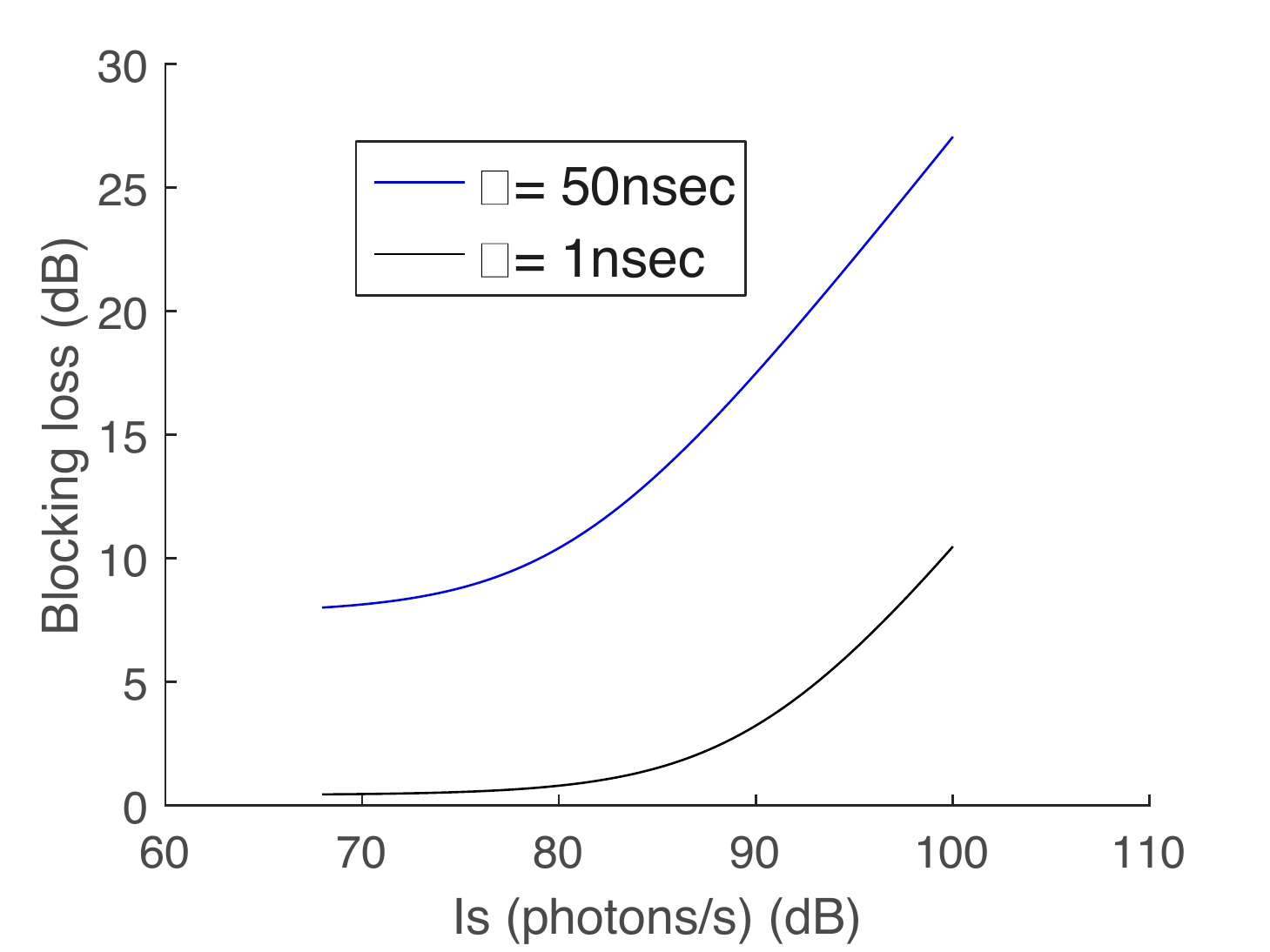}
    \caption{Blocking loss for Mars distances}
    \label{blossMars}
    \end{subfigure}
    \caption{Blocking loss}
    \label{Blockingloss}
\end{figure}

$l_n$ and $l_s$ are the noise and signal photon flux (photons/s) on the detector respectively. $ls=P_{r,ap}\eta_{det}/E$,  
 $ln=P_n/E$,  and $\tau$ denotes the dead time in seconds, i.e., the duration of blocking. 
 $P_n$ is the noise power and $E_{photon}$ is the energy per photon ($hc/\lambda$). $h$ is the Planck’s constant, $c$ is the speed of light, and $\lambda$ is the wavelength. 
 in Fig.~(\ref{Blockingloss}),
$D_t = 0.22m$ and $D_r =4m$ in Fig.~(\ref{blossnormal}), for distances of$ 0.9 -195MKm$ and Fig.~(\ref{blossMars}), for closest to the furthest distance to Mars. As shown as gets further away, blocking loss gets lower, because the number of photon that are received is low. 

\begin{figure}[H]
\centering
\includegraphics[width=0.7\linewidth]{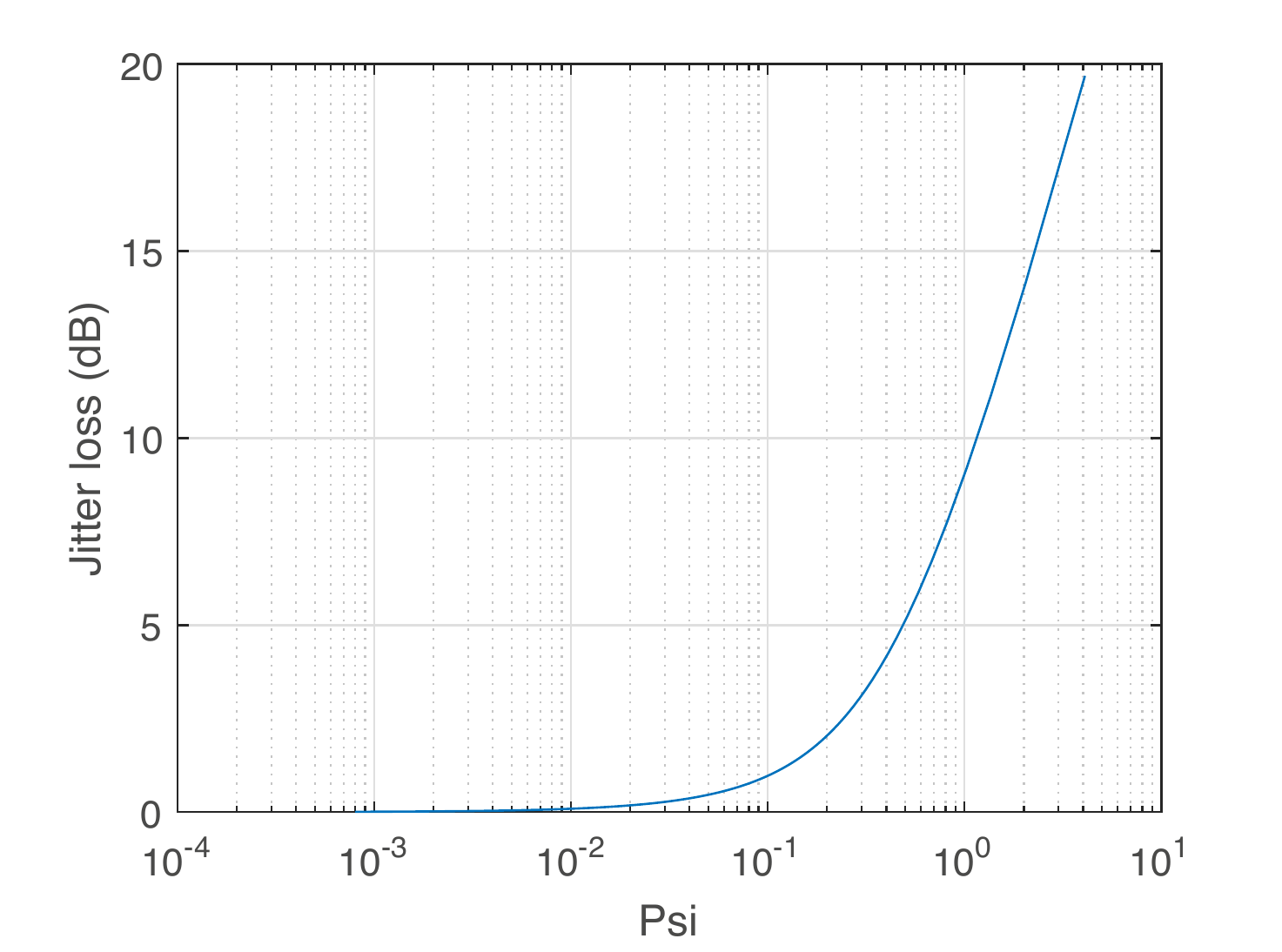}
\caption{Jitter loss}
\label{jitterloss}
\end{figure}

\subsubsection{Detector jitter loss}
In photon-counting detectors, there will be a random delay from the time a photon is incident on the detector to the time an electrical output pulse is generated in response. That random delay, called detector jitter, produces losses since the signaling depends on identifying the pulse position in time analytical expressions for the computation of jitter loss as a function of the normalized jitter variance.
\noindent
\begin{equation}
Lj = 10log10 \left(5{\Psi}+2{\Psi}+1\right)
\end{equation}
\begin{equation}
{\Psi} = \frac{{\sigma}j}{T_{slot}}   \frac{\left(1+\tanh\left({R_{ecc}}-\frac{1}{2}\right) \right)}{1.25^{log_{2}\left(M\right)}} 
\end{equation}
Where $R_{ecc}$ is the coding ratio, $L_{j}$ is jitter detector loss and $\sigma_{j}$ is the standard deviation of jitter.
As shown for $\sigma_{j}/T_{slot}$ less then $1$ jitter loss is negligible but when the ratio is higher then $1$, the losses are sufficiently large, as to typically make the operating point prohibitive~\cite{moision2008Farrjitter}.

\subsection{Required power and coding efficiency}
 
The required power at the receiver is expressed in terms of  coding efficiency and detected power as follows.

\begin{equation}
    {{P}}_{{rq}}{=\ }{{\eta}}_{{coding}}.{{P}}_{{det}}
\end{equation}

Where $\eta_{coding}$ denotes the coding implementation loss and the coding efficiency,  the  gap to the theoretical capacity.
In practice codes only approach the ideal performance. Therefore, code efficiency is the extra power is needed in
order to achieve the ideal performance for a specified BER~\cite{moision2005coded}.

\subsubsection{Noise contribution}

Because of the photon detection nature some  noise sources are accounted, e.g.,  the background noise, dark counts, and leakage power from the transmitter. The total detected noise power is described as follows~\cite{caplan2007laser}:
\begin{equation}
P_n= n_{det}.P_{b}.K_{array}+d^{2}_{detector}.i_{d}. E_{photon}. K_{array}+ {\eta}_{leakage}.P_{r,ap}.{\eta}_{det}
\end{equation}

By working on only one detector $Karray = 1$,  $i_{d}$, measured in electrons per seconds per meter squared, is each detector’s dark rate. $\eta_{leakage}$ is  
the leakage ratio. $Pb$, measured in Watt,  is the background power (to be discussed in more details) per detector, and $d_{detector}$ is the diameter of each detector.

\subsubsection{Background noise} 

The power of the background signal at each detector comes from the  energy diffused from the sky, planets, or
stars which are in the field of view of the receiver :
\begin{equation}
 {{P}}_{{b}}{=\ }{{\Omega}}_{{fov}}{{\eta}}_{{r}}{{A}}_{{r}}({{H}}_{{b,sky}{\ }}+{{H}}_{{b,planet}{\ }}.{{H}}_{{b, stray}})+{{N}}_{{b,star}}.{{\eta}}_{{r}}{{A}}_{{r}}{{B}}_{{r}}
\end{equation}

Here $\Omega_{fov}$ is the field of view of the receiver’s aperture in steradian (Sr). $\Omega_{fov}$ is expresses based on~\cite{Moision2012Designtool,lambert1995laser} as follows
\begin{equation}
  {\Omega}_{fov}=2\pi\left(1-\cos \left(\frac{d_{detector}}{F_{length}} \right)\right)
\end{equation}
Where $F_{length}$ (m) is the focal length of the telescope.
$Ar$ is receiver area in ($m^{2}$), and  $\eta_r$ is the efficiency of the receiver.  $Hb(W/m^{2}/Sr/{\mu}m) $is the background radiance energy density of large angular sources (i.e., planets, sky) and the stray light that scatters into the detector’s field of view. $Hb$ changes in different situations for example $Hb_{sky}$ if it is a terrestrial station on Earth it can get to $54.45 (W/m^{2}/Sr/{\mu}m)$ at a sunny day and at night $Hb_{sky}$ is  $10^{-5} (W/m^{2}/Sr/{\mu}m)$ . 
$Nb (W/m^{2}/{\mu}m)$ is the irradiance energy density of point sources that it might get in the field of view (i.e., stars), which  depend on the wavelength, with that atmospheric losses are also included~\cite{keyITU}.
 $Bf$ is the receive bandpass optical filter width in millimeter  (mm).



\subsection{Numerical analysis for Mars and beyond communication links }
To show how to do some performance analysis, we in the Table.(\ref{parametersvalues}) we provide some values of the system parameters explained through out the previous sections. 
The impact of different parameters on the capacity versus distance is discussed for the set parameters. We set different receiver diameter sizes then,  modulation order,   $T_{slot}$, and vary link distance to make it cover for the change in  Earth-to-Mars distances ($50MKm$ to $401MKm$) as  well as for links at $500MKm$ to $10BKm$ distance form Earth. almost all of the values in the table obtained from~\cite{lyras2019deep}.
\\

\begin{center}
  
   \small
    \begin{longtable}{|c|p{15.25em}|} 
    \caption{Parameters values for Mars link analysis}\\    
     \hline
    \endfirsthead
    \toprule
    \rowcolor[rgb]{ 0,  0,  0} \textcolor[rgb]{ 1,  1,  1}{Deep space Mars Link,  Signal detector } &\textcolor[rgb]{ .137,  .122,  .125}{}  \\
    \midrule
    \rowcolor[rgb]{ 1,  .851,  .4} \multicolumn{1}{|p{20.065em}|}  {Variables}&{INPUTS} \\
    \endhead 
    \midrule
    \rowcolor[rgb]{ .957,  .69,  .514}  \textcolor[rgb]{ .137,  .122,  .125}{Wavelength (nm)} & \textcolor[rgb]{ .137,  .122,  .125}{1550}\\
    \midrule
    \rowcolor[rgb]{ .957,  .69,  .514} \textcolor[rgb]{ .137,  .122,  .125}{Rang (AU)} & \textcolor[rgb]{ .137,  .122,  .125}{0.36/2.68} \\
    \midrule
    \rowcolor[rgb]{ .957,  .69,  .514}  \textcolor[rgb]{ .137,  .122,  .125}{Elevation angle (deg)} &\textcolor[rgb]{ .137,  .122,  .125}{20} \\
    \midrule
    \rowcolor[rgb]{ .957,  .69,  .514}   \textcolor[rgb]{ .137,  .122,  .125}{Transmit power (W) }&\textcolor[rgb]{ .137,  .122,  .125}{4} \\
    \midrule
    \rowcolor[rgb]{ .957,  .69,  .514}  \textcolor[rgb]{ .137,  .122,  .125}{Transmitter diameter (m)} &\textcolor[rgb]{ .137,  .122,  .125}{0.22} \\
    \midrule
    \rowcolor[rgb]{ .957,  .69,  .514}  \textcolor[rgb]{ .137,  .122,  .125}{Transmitter secondary diameter } &\textcolor[rgb]{ .137,  .122,  .125}{0 }\\
    \midrule
    \rowcolor[rgb]{ .957,  .69,  .514} \textcolor[rgb]{ .137,  .122,  .125}{Transmitter efficiency}&  \textcolor[rgb]{ .137,  .122,  .125}{0.6} \\
    \midrule
    \rowcolor[rgb]{ .957,  .69,  .514} \textcolor[rgb]{ .137,  .122,  .125}{Receiver diameter (m)} &\textcolor[rgb]{ .137,  .122,  .125}{4/6/8/10} \\
    \midrule
    \rowcolor[rgb]{ .957,  .69,  .514}  \textcolor[rgb]{ .137,  .122,  .125}{Receiver secondary aperture (m) } & \textcolor[rgb]{ .137,  .122,  .125}{0}\\
    \midrule
    \rowcolor[rgb]{ .957,  .69,  .514} \textcolor[rgb]{ .137,  .122,  .125}{Receiver efficiency} & \textcolor[rgb]{ .137,  .122,  .125}{0.4} \\
    \midrule
    \rowcolor[rgb]{ .957,  .69,  .514}  \textcolor[rgb]{ .137,  .122,  .125}{Receiver quantum  efficiency} & \textcolor[rgb]{ .137,  .122,  .125}{0.5} \\
    \midrule
    \rowcolor[rgb]{ .957,  .69,  .514} \textcolor[rgb]{ .137,  .122,  .125}{Focal length of the receiver (m)} &  \textcolor[rgb]{ .137,  .122,  .125}{16} \\
    \midrule
    \rowcolor[rgb]{ .957,  .69,  .514} \textcolor[rgb]{ .137,  .122,  .125}{Detector diameter (m)} &\textcolor[rgb]{ .137,  .122,  .125}{30e-6} \\
    \midrule
    \rowcolor[rgb]{ .957,  .69,  .514} \textcolor[rgb]{ .137,  .122,  .125}{Optical filter (µm) } &\textcolor[rgb]{ .137,  .122,  .125}{0.2e-3} \\
    \midrule
    \rowcolor[rgb]{ .957,  .69,  .514}\textcolor[rgb]{ .137,  .122,  .125}{Atmospheric efficiency (vertical) } & \textcolor[rgb]{ .137,  .122,  .125}{0.98}  \\
    \midrule
    \rowcolor[rgb]{ .957,  .69,  .514}\textcolor[rgb]{ .137,  .122,  .125}{Link margin (dB)}  & \textcolor[rgb]{ .137,  .122,  .125}{4.0} \\
    \midrule
    \rowcolor[rgb]{ .957,  .69,  .514}  \textcolor[rgb]{ .137,  .122,  .125}{Pointing RMS error value (µrad)} &\textcolor[rgb]{ .137,  .122,  .125}{0.7 } \\
    \midrule
    \rowcolor[rgb]{ .957,  .69,  .514}  \textcolor[rgb]{ .137,  .122,  .125}{Probability level } &\textcolor[rgb]{ .137,  .122,  .125}{1e-14} \\
    \midrule
    \rowcolor[rgb]{ .957,  .69,  .514}  \textcolor[rgb]{ .137,  .122,  .125}{Modulation}& \textcolor[rgb]{ .137,  .122,  .125}{M-PPM} \\
    \midrule
    \rowcolor[rgb]{ .957,  .69,  .514} \textcolor[rgb]{ .137,  .122,  .125}{Modulation numbers }  &\textcolor[rgb]{ .137,  .122,  .125}{16/64/256/1024} \\
    \midrule
    \rowcolor[rgb]{ .957,  .69,  .514}  \textcolor[rgb]{ .137,  .122,  .125}{Slot Time (ns)} &\textcolor[rgb]{ .137,  .122,  .125}{2/0.25} \\
    \midrule
    \rowcolor[rgb]{ .957,  .69,  .514}  \textcolor[rgb]{ .137,  .122,  .125}{Coding ratio} &\textcolor[rgb]{ .137,  .122,  .125}{0.5} \\
  
    \midrule
    \rowcolor[rgb]{ .957,  .69,  .514}  \textcolor[rgb]{ .137,  .122,  .125}{Background noise reduction } &\textcolor[rgb]{ .137,  .122,  .125}{0.5} \\
    \midrule
    \rowcolor[rgb]{ .957,  .69,  .514}  \textcolor[rgb]{ .137,  .122,  .125}{Coding efficiency} &\textcolor[rgb]{ .137,  .122,  .125}{0.8} \\
    \midrule
    \rowcolor[rgb]{ .957,  .69,  .514} \textcolor[rgb]{ .137,  .122,  .125}{Radiance of planets + sky (W/$m^{2}$/Sr/µm) } & \textcolor[rgb]{ .137,  .122,  .125}{85} \\
    \midrule
    \rowcolor[rgb]{ .957,  .69,  .514} \textcolor[rgb]{ .137,  .122,  .125}{Leakage ratio } &\textcolor[rgb]{ .137,  .122,  .125}{0}  \\
    \midrule
    \rowcolor[rgb]{ .957,  .69,  .514} \textcolor[rgb]{ .137,  .122,  .125}{Detector array } &\textcolor[rgb]{ .137,  .122,  .125}{1}  \\
    \midrule
    \rowcolor[rgb]{ .957,  .69,  .514} \textcolor[rgb]{ .137,  .122,  .125}{Detector dark rate (e/s/$m^{2}$)} & \textcolor[rgb]{ .137,  .122,  .125}{ (1e+12)} \\
    \midrule
    \rowcolor[rgb]{ .957,  .69,  .514}  \textcolor[rgb]{ .137,  .122,  .125}{Blocking time (ns)} &\textcolor[rgb]{ .137,  .122,  .125}{50} \\
    \midrule
    \rowcolor[rgb]{ .957,  .69,  .514}  \textcolor[rgb]{ .137,  .122,  .125}{Jitter time (ns)} & \textcolor[rgb]{ .137,  .122,  .125}{240} \\
    \midrule
    \rowcolor[rgb]{ .957,  .69,  .514}   \textcolor[rgb]{ .137,  .122,  .125}{ Atmospheric transmittance } &\textcolor[rgb]{ .137,  .122,  .125}{0.943}\\
    \midrule
    \rowcolor[rgb]{ .957,  .69,  .514} \textcolor[rgb]{ .137,  .122,  .125}{Scintillation loss (dB)} & \textcolor[rgb]{ .137,  .122,  .125}{0.01} \\
    \midrule
    \rowcolor[rgb]{ .957,  .69,  .514}  \textcolor[rgb]{ .137,  .122,  .125}{Pointing loss (dB)} &\textcolor[rgb]{ .137,  .122,  .125}{1.95} \\
    \midrule
    \rowcolor[rgb]{ .957,  .69,  .514}  \textcolor[rgb]{ .137,  .122,  .125}{Cirrus loss (dB)} & \textcolor[rgb]{ .137,  .122,  .125}{0.5} \\\hline
    \end{longtable}
     \label{parametersvalues}
    \end{center}

\begin{figure}[!ht]
\centering
\includegraphics[width=0.7\linewidth]{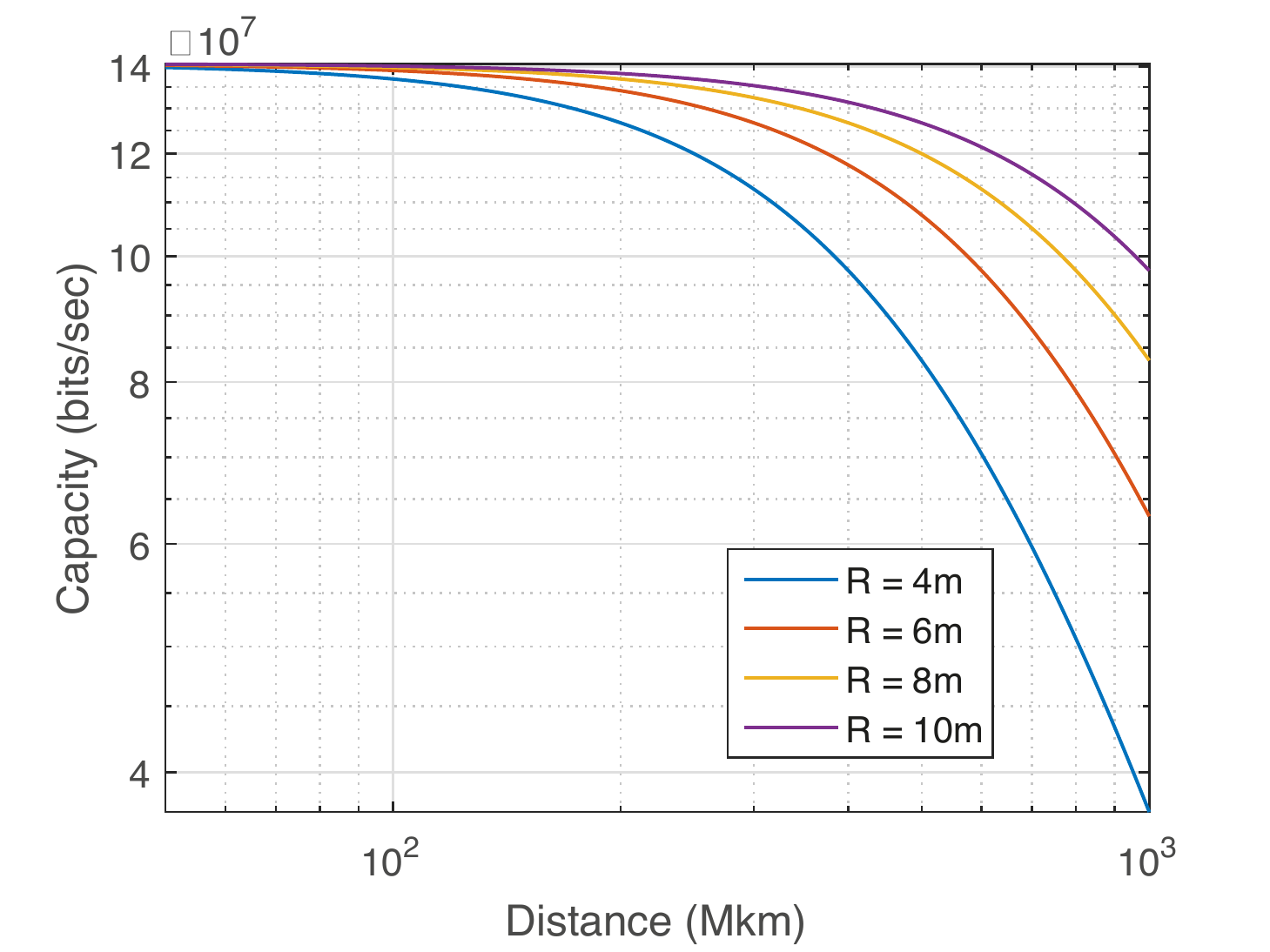}
\caption{Capacity vs distance for different receiver diameters at $M=16$}
\label{capacityvsDistance}
\end{figure}

The result obtained in Fig.~(\ref{capacityvsDistance}), calculated capacity at $ M = 16$ and $T_{slot} = 0.25ns $ shows that the bigger receiver diameter, $R$ the higher capacity is achieved also in the range of Mars distances it seems that the slop for diameters of $4m$ is significantly higher then $D_r = 10$.

\begin{figure}[!ht]
\centering
\includegraphics[width=0.7\linewidth]{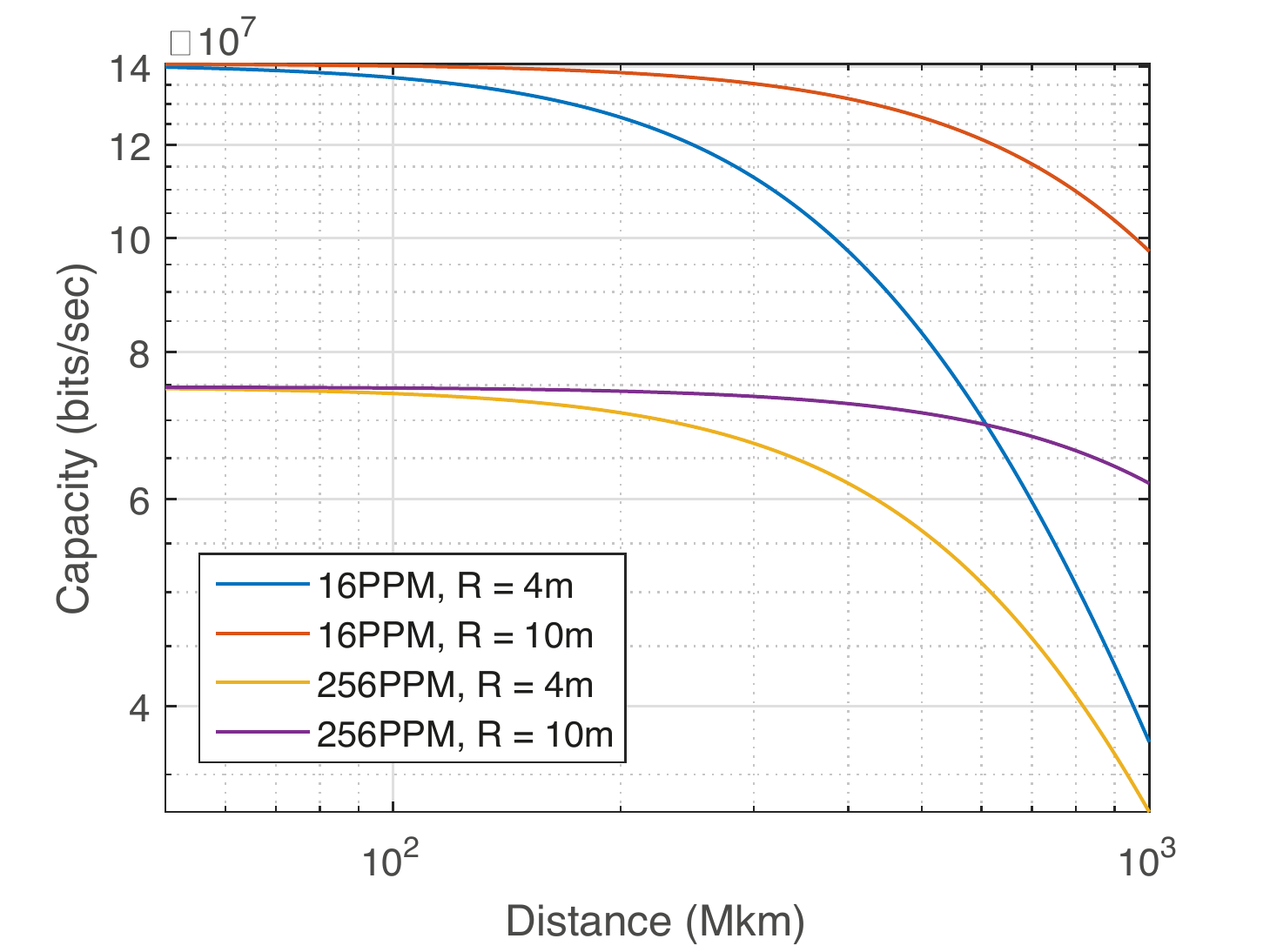}
\caption{Capacity vs distance for various receiver diameters and modulation orders}
\label{capacityfordifferentM}
\end{figure}

In Fig.~(\ref{capacityfordifferentM}), calculating the capacity for$  D_{r} = 4, 10 m$ and $T_{slot}= 0.25ns$  but one for M= 16 and another at M = 256 and as shown for M=16 gives higher capacity then M = 256 in this rang of distances.\\
\begin{figure}[!t]
\centering
\includegraphics[width=0.7\linewidth]{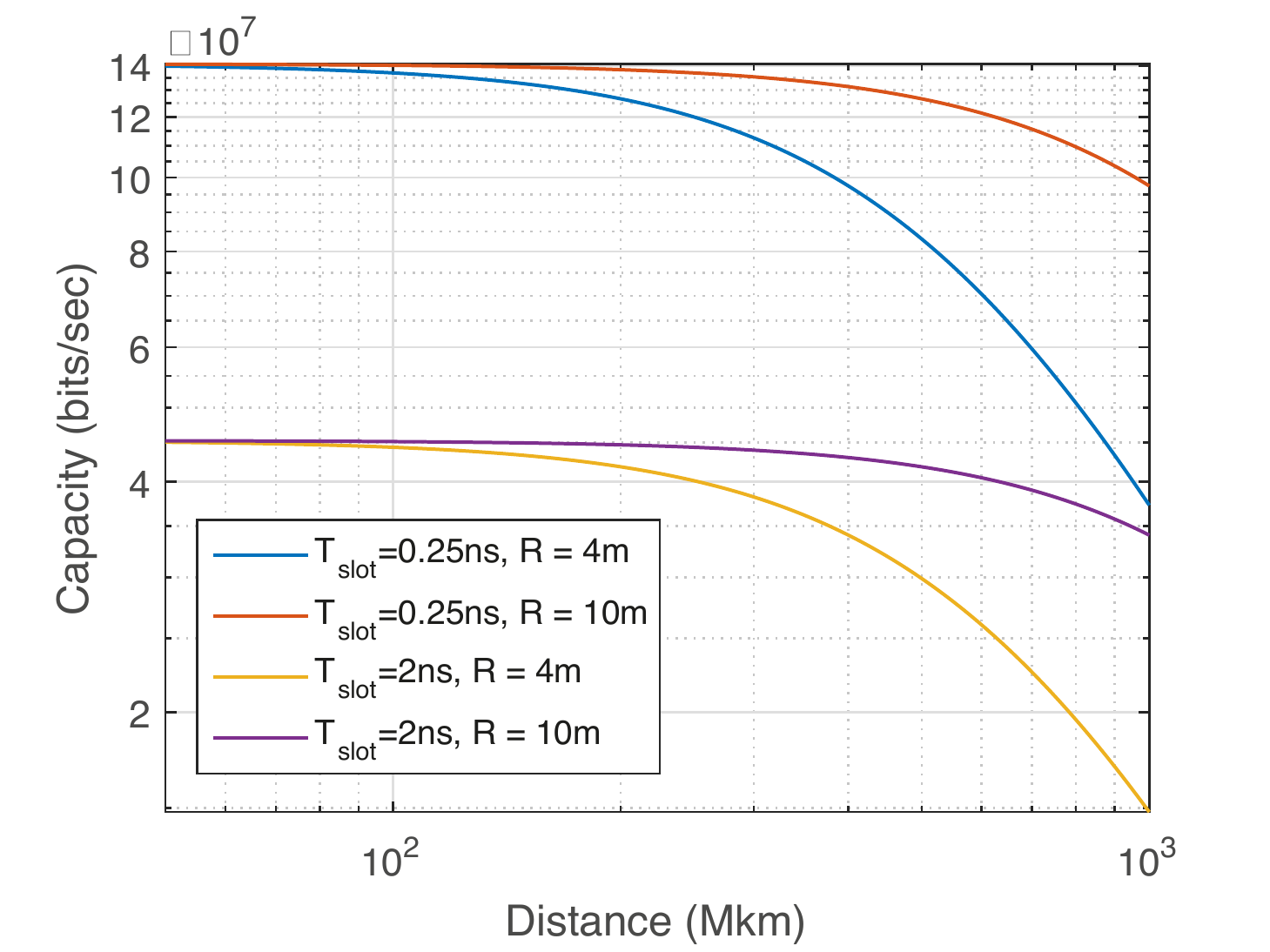}
\caption{Capacity vs distance for various receiver diameters and time slots at $M=16$}
\label{capacitydifferentT}
\end{figure}

In Fig.~(\ref{capacitydifferentT}), calculating the capacity for $D_{r} = 4,10 m$, M= 16, and both $T_{slot}$ = 2ns and  $T_{slot}$ = 0.25ns. The lower $T_{slot}$ the higher the capacity for the given range of distances, as the figure explains.

\begin{figure}[!t]
\centering
\includegraphics[width=0.7\linewidth]{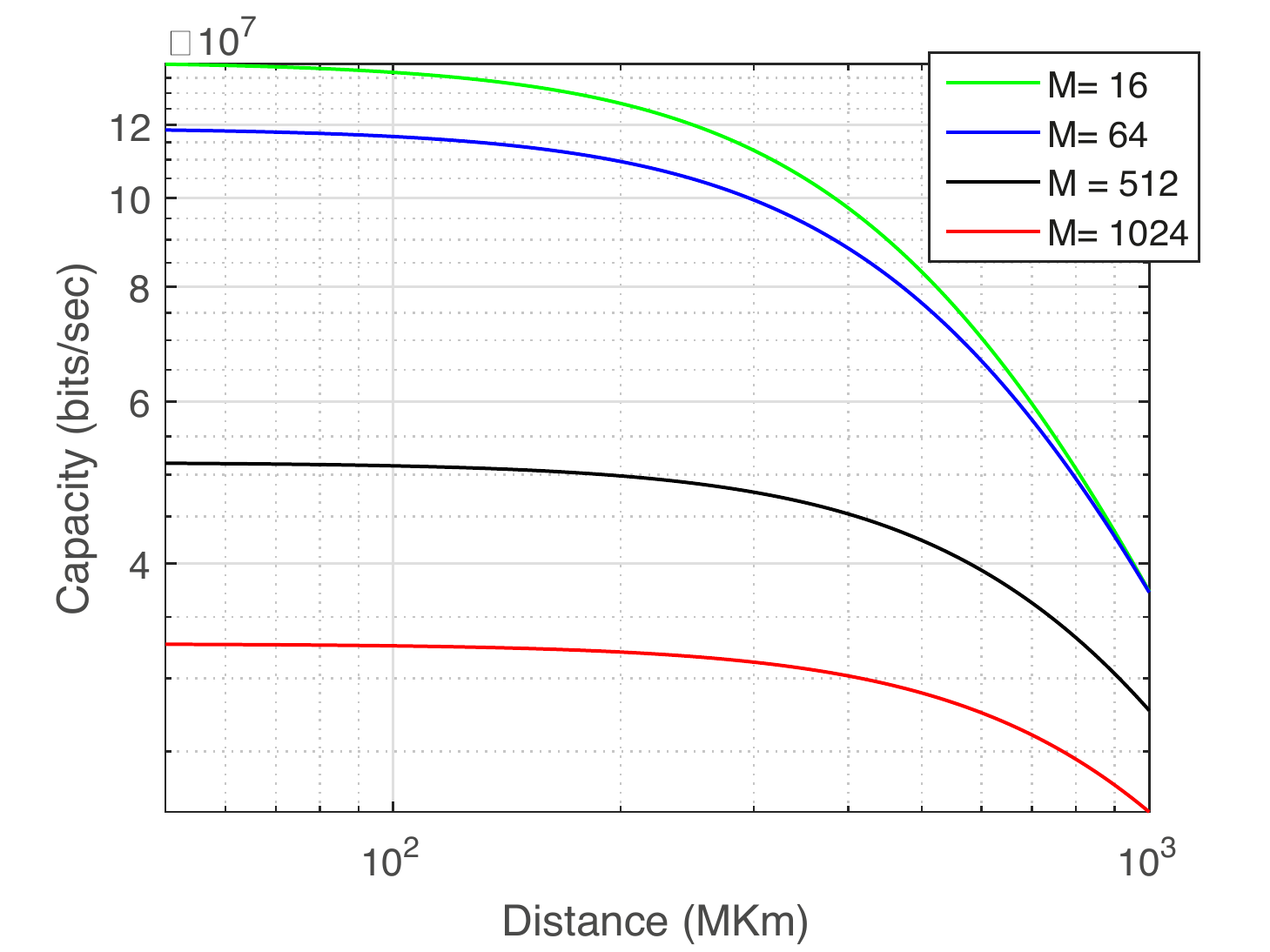}
\caption{Capacity vs distance for various modulation orders for Mars distances with $D_{r}= 4m$.}
\label{MarsD4}
\end{figure}

\begin{figure}[!ht]
\centering
\includegraphics[width=0.7\linewidth]{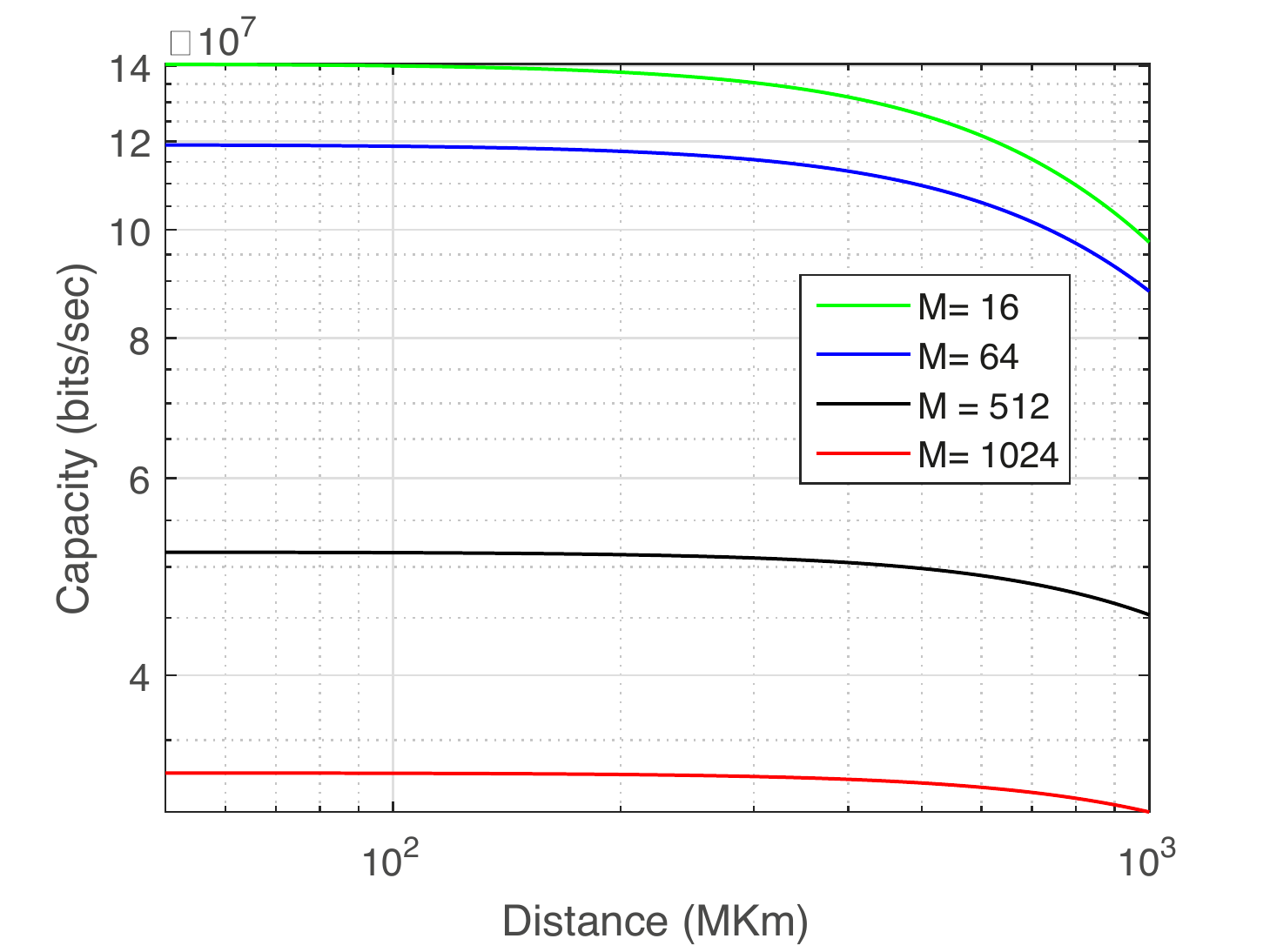}
\caption{Capacity vs distance for various modulation orders for Mars distances with  $D_{r} = 10m$}
\label{MarsD10}
\end{figure}

\begin{figure}[!ht]
\centering
\includegraphics[width=0.7\linewidth]{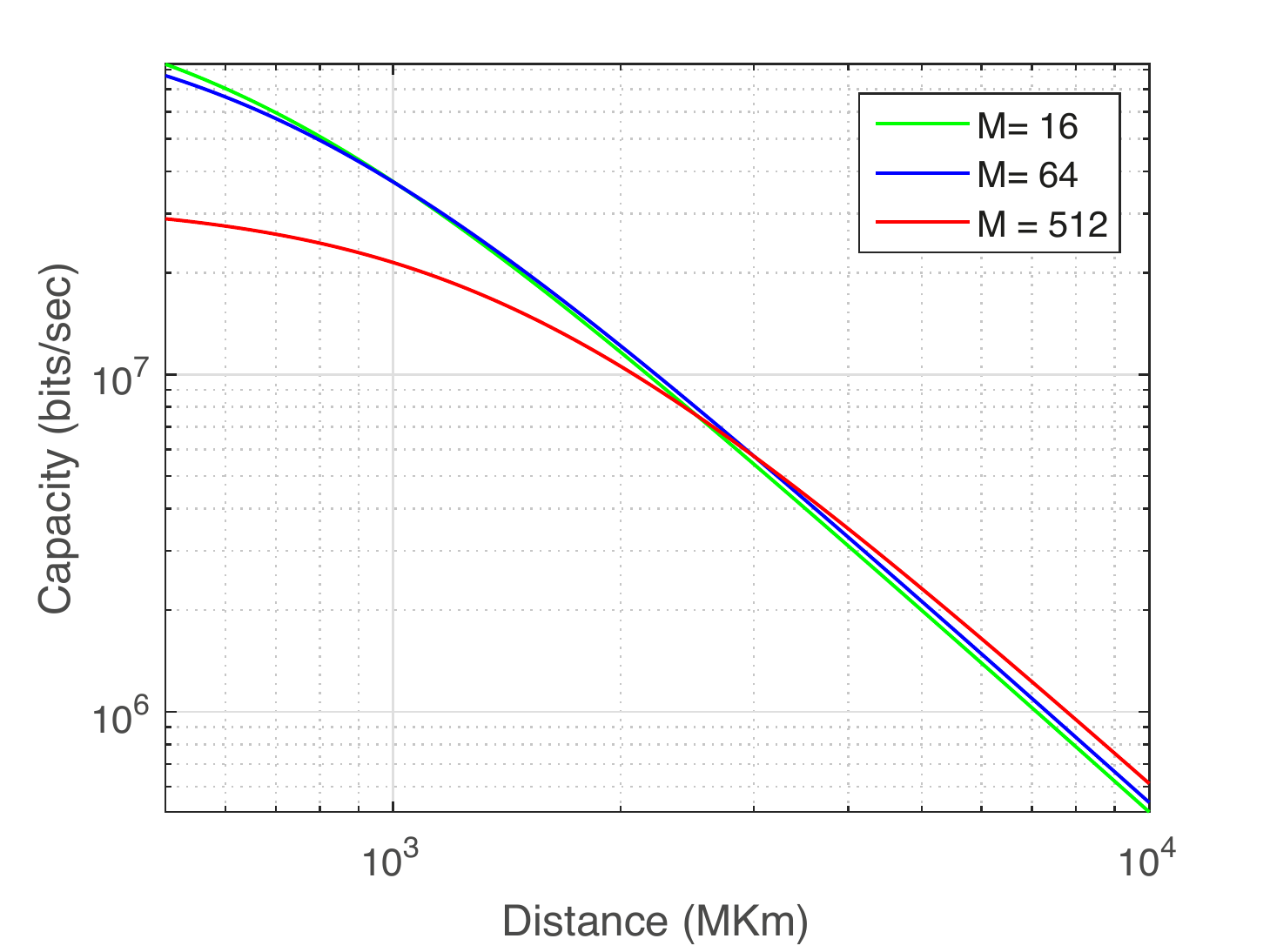}
\caption{Capacity vs distance for various modulation orders for 500Mkm to 10Bkm away from Earth  with $D_{r}= 4m$  }
\label{longDisD4}
\end{figure}

\begin{figure}[!ht]
\centering
\includegraphics[width=0.7\linewidth]{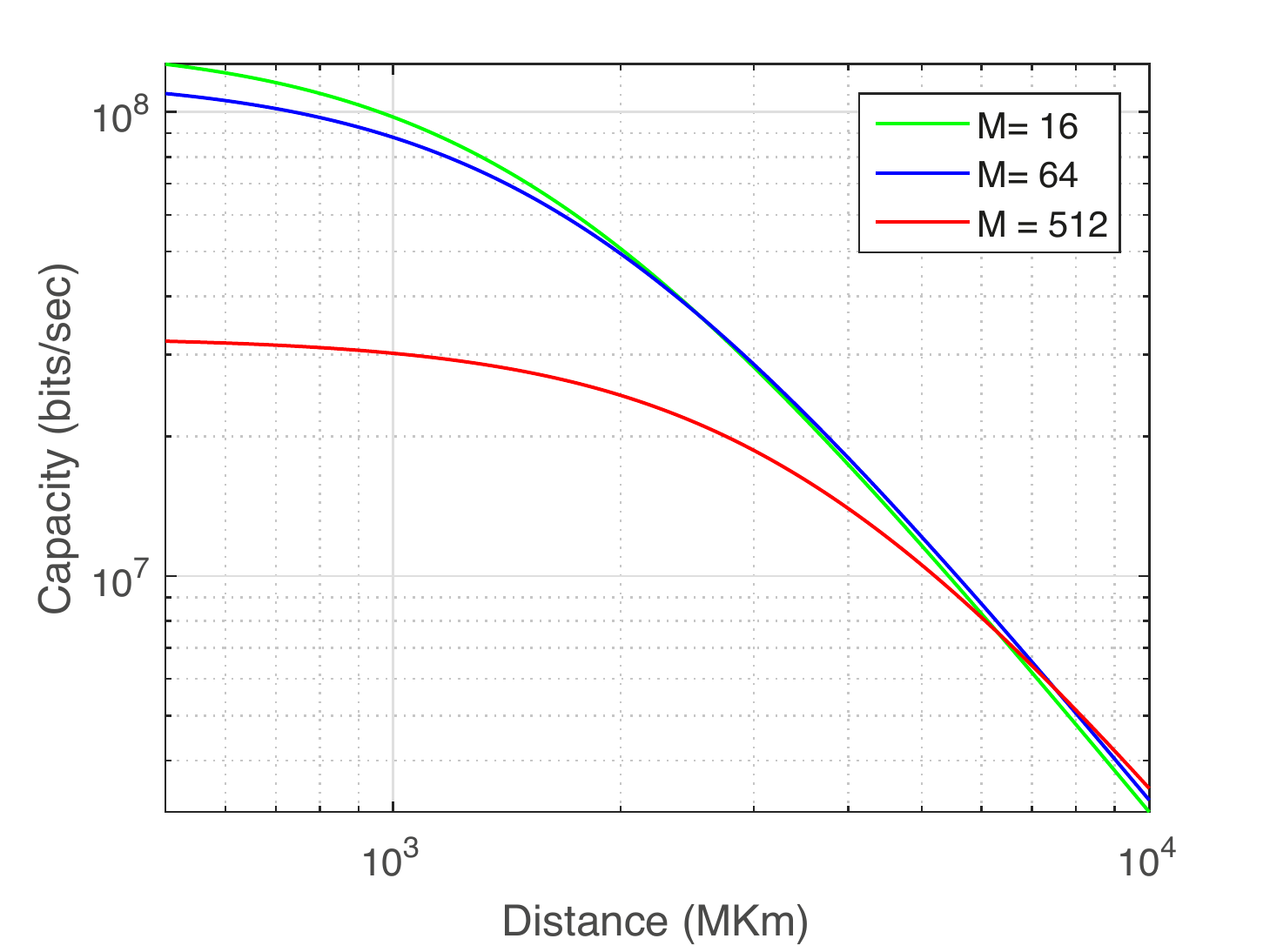}
\caption{Capacity vs distance for various modulation orders for 500Mkm to 10Bkm away from Earth  with $D_{r}= 10m$ }
\label{longDisD10}
\end{figure}

 In Fig.~(\ref{MarsD4}) and .~(\ref{MarsD10}), is shown that $M=16$ gives the best capacity possible for the link for $D_{r} = 4m$ and $D_{r} = 10m$, but for the later is shown that the $M=16$ capacity line has lower slop along Mars distances which indicates that bigger receiver diameters gives more satiable capacity along the Mars distance.\\
As going further away for Mars distances, we get from $500Mkm$ to $10Bkm$ from Earth does the response be the same?, Calculating the capacity for different M orders in this range of distances and the response shown in Fig.~(\ref{longDisD4}), in the case of $D_{r} = 4m$ and Fig.~(\ref{longDisD10}), for $D_{r} = 10m$.\\
 
 Fig.~(\ref{MarsD4}) and  Fig.~(\ref{longDisD4})  we show achieved capacity of PPM system   versus Mars  and beyond Mars distances respectively, antenna diameter for transmitter and receiver are  $0.22m$  and $4m$ respectively.
The figure indicates how crucial the increase in DS communication distance is in determining the PPM system parameters and hence performance. As the modulation order decreases,  the capacity of a given system improves by a significant margin until a certain distance. However 
as communication path takes a longer distance (in our settings close to $2000MKm$), the trend is the opposite. 
That is intuitive as increasing   modulation order implies a reduction in the transmission rate and  bit error rate (BER)   as a consequence which becomes more crucial as distances increase (which results in a decreasing in the received power). In other words, the distance may impose us to increase the modulation order and hence the transmission rate to improve the system performance and maintain the highest capacity possible for a given channel model, detector, and other system parameters like receiver antenna diameter,  and transmission power.

 In the case of  $D_{r} = 4m$ in Fig.~(\ref{longDisD4}), starting from roughly $100Mbps$ which is accomplished by $M = 16$, if $M = 64$ is chosen, a capacity as high as $5Mbps$ is not used. As the distance gets further away at $1000Mkm$ $M = 64$ will get the highest capacity as it cross the $M = 16$ line, and at $3000Mkm$ an beyond $M = 1024$ gets the highest capacity. If  there is not a switch to $M = 64$ when it crosses $M=16$ and $M = 1024$ as it crosses $M = 64$, a capacity of $~0.5Mbps$ is not utilized. For Fig.~(\ref{longDisD10}), $D_{r} = 10m$ beginning with $~125Mpbs$ for $M=16$, if  order $64$ or $1024$ is chosen, $(15 - 90)Mbps$ capacity is not used. At  $250Mkm$ $M = 64$ cross $M = 16$ and gives the highest capacity and takes over, and the different between the capacity for $M = 64$ to $M = 16$ again reaches to $0.5Mbps$. in distance of $7500Mkm$ $M = 1024$ gives the highest capacity of the link. Here if the link settled on one $M$ order along this range of distances, amount of capacity as big as $0.5Mbps$ is not utilized, this amount of data rate could be significant, which  suggests using different Modulation number at different distances to Mars and beyond or along a missions.

\subsection{\acrshort{ccsds} standard for selecting link parameters}
After analysing performance of the link with different parameters values there is a  methodology based on \acrshort{ccsds} aims at selecting parameters that maximize the data rate of the link and avoids tedious coded BER evaluations. Especially in case of a single photon detector or an array.
It is mainly based on first considering all the  combination of the  link parameters which are the available PPM order, $T_{slot}$, $R_{ecc}$ coding ratio, and the computing the data rate with \acrshort{scppm} coding for all the combination, and computing the received power by link analysis and the detected noise power,  and then commuting the soft capacity of each power received calculated and noise power, based on that, selecting the best $R_{ecc}$ , $M$, and $T_{slot}$ that gives the highest data rate. Note that achieved data rate will be smaller than the capacity. However, if $R_{ecc}$  was free to take any value, then the capacity and the achieved data rate will be closer.

\section{Discussion}

Starting with how to select $T_{slot}$, $R_{ecc}$ , and the $M$ order for getting specific data rate and conclude the power required, seeing the behavior of \acrshort{ppm} with different $M$ order verse power, the lower the power the higher order is respired for getting high capacity. Noise has some power threshold, if its passed, noise will not have an affect on the link performance. Getting more in details in the primary effects on the link which its major loss would be due to the long distance operation. Detection is effected with blocking and jitter loss, It would be low for low data rate operation, because not high number of photons are detected which is the case for DS communication. And further more, in numerical analysis as shown in Mars distances with the parameters used, $M=16$ is the dominated order to get the highest capacity. But that would change at different parameters values. For distances starting form $500Mkm$ to $10BKm$ we got capacity in the range of $(125 - 3.5) Mbps$ for receiver's diameter of $10m$, and $(100 - 0.6) Mpbs$ for receiver's diameter of $4m$, and a loss of $0.5Mbps$ if $M=16$ is chosen along the whole distance.

\chapter{Conclusion}

In this thesis, by performing link analysis in Mars distances using PPM, exploring FSO technologies in all network layers, and analyzing OAM propagation in DS, along with discussions on how to mitigate the advances' challenges, we demonstrated the effectiveness and superiority of FSO compared to RF's expected performance for DS communications. According to the PPM link analysis, the order of modulation that achieves the maximum capacity varies at different distances. We challenged the PPM scheme and the PC detection with other techniques for obtaining a higher capacity limit. We focused our attention on the key features and protocols of the DS network layers architecture and related possible enhancements that may lead to a interplanetary network. Further, we investigated how promising technologies would impact DS communications and pointed out the struggles and proposed solutions for accommodating OAM in DS environments which require further investigation.


\cleardoublepage
\addcontentsline{toc}{chapter}{References}
\renewcommand{\bibname}{References}
\printbibliography

\cleardoublepage
  \break
  \pagenumbering{arabic}
\renewcommand*{\thepage}{A\arabic{page}}
\appendix

\chapter{Appendix}

\section{Power received against the capacity for different $M$ orders}
\label{mappingcode}

\begin{lstlisting}[language=MATLAB]

lammbda = 1550*(10^-9) ;

E = 1.28*(10^(-19)) ;
Pr1 = -160:1:1 ;
Pr = 10.^(Pr1/10);

Pn = 2.1*(10^(-14)) ;
M = 16;
Tslot = 1*(10^-9);

term1 = (1/(log(2)*E));
% case 0
 term11 = 1 ./((Pr./log(8))+((Pn*2)/(8-1))...
     +(((Pr.^2)*(8*Tslot))/(log(8)*E)));
 C = term1.*((Pr.^2).*term11);

% case 1 
 term2 = 1 ./((Pr./log(M))+((Pn*2)/(M-1))+...
     (((Pr.^2)*(M*Tslot))/(log(M)*E)));
 C1 = term1.*((Pr.^2).*term2);
 %case 2
 term22 = 1 ./((Pr./log(64))+((Pn*2)/(64-1))...
     +(((Pr.^2)*(64*Tslot))/(log(64)*E)));
 C2 = term1.*((Pr.^2).*term22);
% case 3
 term23 = 1 ./((Pr./log(256))+((Pn*2)/(256-1))...
     +(((Pr.^2)*(256*Tslot))/(log(256)*E)));
 C3 = term1.*((Pr.^2).*term23);

% case 4 

 term24 = 1 ./((Pr./(log(512)))+((Pn*2)/(512-1))...
     +(((Pr.^2)*((512)*Tslot))/(log(512)*E)));
 C4 = term1.*((Pr.^2).*term24);
 
 % case 5 

 term25 = 1 ./((Pr./(log(1024)))+((Pn*2)/(1024-1))...
     +(((Pr.^2)*((1024)*Tslot))/(log(1024)*E)));
 C5 = term1.*((Pr.^2).*term25);
 
  % case 6 

 term26 = 1 ./((Pr./(log(2048)))+((Pn*2)/(2048-1))...
     +(((Pr.^2)*((2048)*Tslot))/(log(2048)*E)));
 C6 = term1.*((Pr.^2).*term26);
 

 figure
  
 loglog(Pr,C,'b','LineWidth',1.15)
 hold on
 loglog(Pr,C1,'r','LineWidth',1.15)
 loglog(Pr,C2,'k','LineWidth',1.15)
 loglog(Pr,C3,'g','LineWidth',1.15)
 loglog(Pr,C4,'c','LineWidth',1.15)
 loglog(Pr,C5,'m','LineWidth',1.15)
 loglog(Pr,C6,'y','LineWidth',1.15)
 
 title('Capacity vs Power')
      xlabel('power (W)')
      ylabel('Capacity (Bits/sec)')
      legend( 'M =8 ','M =16 ','M = 64','M = 256','M = 512', 'M = 1024','M = 2048')
grid on




\end{lstlisting}

\section{Power received against the capacity for different noise power values for $M=16$}
\label{mappingcode}

\begin{lstlisting}[language=MATLAB]

lammbda = 1550*(10^-9) ;

E = 1.28*(10^(-19)) ;
Pr1 = -160:1:1 ;
Pr = 10.^(Pr1/10);
Pn = (10^(-13)) ;
M = 16;
Tslot = 0.25*(10^-9);
(1.38*(10^(-13)))

term1 = (1/(log(2)*E));

% case 1 
 term2 = 1 ./((Pr./log(M))+((Pn*2)/(M-1))+(((Pr.^2)*(M*Tslot))...
     /(log(M)*E)));
 C1 = term1.*((Pr.^2).*term2);
 %case 2
 term22 = 1 ./((Pr./log(M))+((((10^(-12)))*2)/(M-1))...
     +(((Pr.^2)*(M*Tslot))/(log(M)*E)));
 C2 = term1.*((Pr.^2).*term22);
% case 3
 term23 = 1 ./((Pr./log(M))+((((10^(-11)))*2)/(M-1))...
     +(((Pr.^2)*(M*Tslot))/(log(M)*E)));
 C3 = term1.*((Pr.^2).*term23);

% case 4 

 term24 = 1 ./((Pr./(log(M)))+((((10^(-10)))*2)/(M-1))...
     +(((Pr.^2)*((M)*Tslot))/(log(M)*E)));
 C4 = term1.*((Pr.^2).*term24);
 
 % case 5 

 term25 = 1 ./((Pr./(log(M)))+((((10^(-9)))*2)/(M-1))...
     +(((Pr.^2)*((M)*Tslot))/(log(M)*E)));
 C5 = term1.*((Pr.^2).*term25);
 
  % case 6 

 term26 = 1 ./((Pr./(log(M)))+((((10^(-8)))*2)/(M-1))...
     +(((Pr.^2)*((M)*Tslot))/(log(M)*E)));
 C6 = term1.*((Pr.^2).*term26);
 

 figure


 loglog(Pr,C1,'r','LineWidth',1.15)
  hold on
 loglog(Pr,C2,'k','LineWidth',1.15)
 loglog(Pr,C3,'g','LineWidth',1.15)
 %loglog(Pr,C4,'c')
 loglog(Pr,C5,'m','LineWidth',1.15)
% loglog(Pr,C6,'y')
 
 title('Capacity vs Power M = 16')
      xlabel('power (W)')
      ylabel('Capacity (Bits/sec)')
      
        legend( 'Pn = 10^-13 ','Pn = 10^-12 ','Pn = 10^-11','Pn = 10^-9 ')
    





\end{lstlisting}

\section{Maximum capacity at different noise power values}
\label{mappingcode}

\begin{lstlisting}[language=MATLAB]

lammbda = 1550*(10^-9) ;

E = 1.28*(10^(-19)) ;
Pn = (10^(-14)) ;
Pn2 = (10^(-12)) ;
Pn3 = (10^(-10)) ;
Pn4 = (10^(-8)) ;
Tslot = 0.25*(10^-9);
M=[4 , 8 ,16 ,32 ,64 ,128,256 ,512 ,1024 ,2048 ,4096 , 8192 ,16384 ,...
    32768 , 65536 ,131072 ,262144 , 524288 , 1048576];
C1 = [] ;
C2 = [] ;
C3 = [] ;
C4 = [] ;
%*******************************************************
%Pn 1 
j = 1 ;
for  Pr1 = -160:1:1
 for i = 1:1:19  ;
Pr = 10.^(Pr1/10); 
 term1 = (1/(log(2)*E));
% case 1 
 term2 = 1./((Pr./log(M(i)))+((Pn*2)./(M(i)-1))...
     +(((Pr.^2).*(M(i)*Tslot))./(log(M(i))*E)));
 C1(i) = term1.*((Pr.^2).*term2);
 end
 [max_val max_ind]=max(C1)
 argmax1(j)=M(max_ind)
 j = j+1
end
%********************************************************
% Pn 2
j = 1 ;
for  Pr1 = -160:1:1
 for i = 1:1:19  ;
Pr = 10.^(Pr1/10); 
 term1 = (1/(log(2)*E));
% case 1 
 term2 = 1./((Pr./log(M(i)))+((Pn2*2)./(M(i)-1))...
     +(((Pr.^2).*(M(i)*Tslot))./(log(M(i))*E)));
 C2(i) = term1.*((Pr.^2).*term2);
 end
 [max_val max_ind]=max(C2)
 argmax2(j)=M(max_ind)
 j = j+1
end
%********************************************************
%Pn 3
j = 1 ;
for  Pr1 = -160:1:1
 for i = 1:1:19  ;
Pr = 10.^(Pr1/10); 
 term1 = (1/(log(2)*E));
% case 1 
 term2 = 1./((Pr./log(M(i)))+((Pn3*2)./(M(i)-1))...
     +(((Pr.^2).*(M(i)*Tslot))./(log(M(i))*E)));
 C3(i) = term1.*((Pr.^2).*term2);
 end
 [max_val max_ind]=max(C3)
 argmax3(j)=M(max_ind)
 j = j+1
end

%********************************************************
%Pn 3
j = 1 ;
for  Pr1 = -160:1:1
 for i = 1:1:19  ;
Pr = 10.^(Pr1/10); 
 term1 = (1/(log(2)*E));
% case 1 
 term2 = 1./((Pr./log(M(i)))+((Pn4*2)./(M(i)-1))...
     +(((Pr.^2).*(M(i)*Tslot))./(log(M(i))*E)));
 C4(i) = term1.*((Pr.^2).*term2);
 end
 [max_val max_ind]=max(C4)
 argmax4(j)=M(max_ind)
 j = j+1
end

%****************************************************
%Plots 

Pr1 = -160:1:1 ;
Pr = 10.^(Pr1/10);

figure
loglog(Pr,argmax1,'r','LineWidth',1.15)
 hold on
 loglog(Pr,argmax2,'g','LineWidth',1.15)
 loglog(Pr,argmax3,'b','LineWidth',1.15)
 loglog(Pr,argmax4,'k','LineWidth',1.15)

 title('Power received vs optimal PPM order ')
      xlabel('power (W)')
      ylabel('M Order')
     legend('Pn = 10^-14','Pn = 10^-12','Pn = 10^-10','Pn = 10^-9')





\end{lstlisting}

 \section{Blocking loss}
\label{mappingcode}

\begin{lstlisting}[language=MATLAB]
%power recived

lammbda = 1550*10^-9 ;
Z = 50 :1: 401 ;
Z = Z.*(10e+9);
E = 1.28*10^(-19) ;
Dt = 0.22 ;
Dr = 4 ;
pt = 4;
Gt = ((pi*Dt)/lammbda)^2;
Gr = ((pi*Dr)/lammbda)^2;
% free space loss 
Lf = (lammbda ./(4*pi.*Z)).^2 ;
% atmospheric transmitenc loss 
La = 0.943 ;
%cirrus transmittance factor 
Lc = 1.12 ;
% the scintillation loss factor 
Ls = 1 ; 
%pointing loss factor 
Lp = 0.6384 ;
phit = 0.6 ;
phir = 0.4 ;

 pr = pt * Gt * Gr * Lf * La * Lc * Lp * phit *phir ;
%******************************************************************

% Blocking loss 
taw1 = 50 * (10^-9) ;
taw2 = 1 * (10^-9) ;
ln = 10^8 ; 
ls = ((pr .* 0.5)./E ) ;
l= ln + ls ; 
Lb1 = 1./(1+(taw1.*l));
Lb2 = 1./(1+(taw2.*l));
Lb1 = 10 * log10(Lb1);
Lb2 = 10 * log10(Lb2);
Lb1 = Lb1 * (-1) ;
Lb2 = Lb2 * (-1) ;
ls = 10 * log10(ls);

 
 figure
hold on
plot(ls,Lb1,'b')
plot(ls,Lb2,'k')
     title('Blocking loss')
     xlabel('Is (photons/s) (dB)');
     ylabel('Blocking loss (dB)');
     legend('t=50nsec ','t=1nsec' );
     
 
 



\end{lstlisting}

 \section{Jitter loss}
\label{mappingcode}

\begin{lstlisting}[language=MATLAB]

segma = 1;
Ts = 0.1:0.1:512 ;
Recc = 0.5; 
M = 16 ;
Psi = (segma ./Ts)*((1+ (tanh(Recc - 0.5)))./(1.25^4)); 

lif = (5*(Psi.^2))+(2.*Psi)+1 ;
Li = 10*log10(lif);

semilogx(Psi , Li);
title('jitter loss')
xlabel('psi');
     ylabel('jitter loss (dB)');
\end{lstlisting}

 \section{Data rate at different receiver diameter size}
\label{mappingcode}

\begin{lstlisting}[language=MATLAB]

M=256;
Z = 50:1: 500 ;
Z = Z.*(10e+9);
%for slot Time of 2ns ;

Tslot = 2*10^-9 ;

% case 1 
Pr = PowerReceived( Tslot ,4 ,M ,Z );
Pn =  NoiesPowerRececived( 4,Z );
term2 = 1 ./((Pr./log(M))+((Pn*2)/(M-1))+((Pr.^2)*(M*Tslot))/(log(M)*E));
C1 = term1.*((Pr.^2).*term2);
% case 2
Pr = PowerReceived( Tslot , 6 ,M,Z );
Pn =  NoiesPowerRececived( 6 ,Z );
term22 = 1 ./((Pr./log(M))+((Pn*2)/(M-1))+((Pr.^2)*(M*Tslot))/(log(M)*E));
C2 = term1.*((Pr.^2).*term22);
% case 3
Pr = PowerReceived(Tslot , 8 ,M, Z);
Pn =  NoiesPowerRececived( 8,Z );
term23 = 1 ./((Pr./log(M))+((Pn*2)/(M-1))+((Pr.^2)*(M*Tslot))/(log(M)*E));
C3 = term1.*((Pr.^2).*term23);

% case 4 
Pr = PowerReceived( Tslot ,10 ,M ,Z);
Pn =  NoiesPowerRececived( 10 ,Z );
term24 = 1 ./((Pr./(log(M)))+((Pn*2)/(M-1))+((Pr.^2)*((M)*Tslot))/(log(M)*E));
C4 = term1.*((Pr.^2).*term24);
close all

figure

loglog(Z,C1,Z,C2,Z,C3,Z,C4 ,'LineWidth',1.15);
     title('Capacity vs Distance for')
     xlabel('Distance(m)')
     ylabel('Capacity (Bits/sec)')
    title ('Data rate versus Hb = 85')
     legend('R =4m ','R = 6m','R = 8m' , 'R = 10m')
     grid on


\end{lstlisting}

 \section{Power received}
\label{mappingcode}

\begin{lstlisting}[language=MATLAB]
function [ Pr] = PowerReceived( Ts,Dr,M ,Z)
%power recived

lammbda = 1550*10^-9 ;
Dt = 0.22 ;
pt = 4;
Gt = 2*((pi*Dt)/lammbda)^2;
Gr = ((pi*Dr)/lammbda)^2;
% free space loss 
Lf = (lammbda ./(4*pi.*Z)).^2 ;
% atmospheric transmittance loss 
La = 0.943 ;
% cirrus transmittance factor 
Lc = 1.12 ;
% the scintillation loss factor 
Ls = 1 ; 
% pointing loss factor 
Lp = 1.5667 ;
% margin
Mar = 2.51 ; 
phit = 0.6 ;
phir = 0.4 ;

 prap = pt * Gt * Gr * Lf * La * Lc * Ls* Lp * phit *phir *Mar ;
%******************************************************************


phid = 0.5 ;
k = 1 ; 
DetectorD = 30*(10^-6 );
td = 10^12 ; 
E = 1.28*10^(-19) ;
phiLeakeag = 0 ;
Flength = 16 ;
Ar = pi*(Dr/2)^2 ; 
Bf= 0.2*(10^-9) ; 
%**************************************************************
H = 15 ; 
Omega = 2*pi*(1 -(cos(DetectorD/(2*Flength))));
Ns = 1.94462*(10^-9);
Pp = (H*Omega*phir*Ar*Bf)+ (Ns*phir*Ar*Bf);
Pn = (phid*Pp*k)+((DetectorD^2)*td*E*k)+(phiLeakeag*prap*phid);
Lb= Blocklossvalue(prap, Pn );
Lj= jitterlossvalue( Ts ,M );
Pd= prap.*phid.*Lb.*Lj ;
phic= 0.8;
Pr=Pd.*phic;
Summ = Lb + Lj
end

\end{lstlisting}

 \section{Noise power received}
\label{mappingcode}

\begin{lstlisting}[language=MATLAB]
function [ Pn ] = NoiesPowerRececived( Dr ,Z )
%UNTITLED4 Summary of this function goes here
%   Detailed explanation goes here

lammbda = 1550*10^-9 ;
Dt = 0.22 ;
pt = 4;
Gt = 2*((pi*Dt)/lammbda)^2;
Gr = ((pi*Dr)/lammbda)^2;
% free space loss 
Lf = (lammbda ./(4*pi.*Z)).^2 ;
% atmospheric transmittance loss 
La = 0.943 ;
% cirrus transmittance factor
Lc = 1.12 ;
% the scintillation loss factor 
Ls = 1 ; 
% pointing loss factor 
Lp = 1.5667 ;
% margin
Mar = 2.51 ;
phit = 0.6 ;
phir = 0.4 ;
summ = La+Lc+Ls+Lp
 pr = pt * Gt * Gr * Lf * La * Lc * Lp * phit *phir *Mar  ;
%******************************************************************


phid = 0.5 ;
k = 1 ; 
DetectorD = 30*(10^-6 );
td = 10^12 ; 
E = 1.28*10^(-19) ;
phiLeakeag = 0 ;
Flength = 16 ;
Ar = pi*(Dr/2)^2 ; 
Bf= 0.2*(10^-9) ; 
%**************************************************************
H = 15 ; 
Omega = 2*pi*(1 -(cos(DetectorD/(2*Flength))));
Ns = 1.94462*(10^-9);
Pp = (H*Omega*phir*Ar*Bf)+ (Ns*phir*Ar*Bf);
Pn = (phid*Pp*k)+((DetectorD^2)*td*E*k)+(phiLeakeag*pr*phid);

end

\end{lstlisting}

 \section{Blocking loss value}
\label{mappingcode}

\begin{lstlisting}[language=MATLAB]
function [ Lb1 ] = Blocklossvalue(pr, pn )

E = 1.28*10^(-19)
taw1 = 50 * (10^-9) ;
ln = (pn/E) ; 
ls = ((pr*0.5)/E ) ;
l= ln + ls ; 
Lb1 = 1./(1+(taw1.*l));
end

\end{lstlisting}

 \section{Jitter loss value}
\label{mappingcode}

\begin{lstlisting}[language=MATLAB]
function [ lif ] = jitterlossvalue( Ts ,M )

segma = 240*10^(-12);
Recc = 0.5; 
Psi = (segma/(Ts))*((1+ (tanh(Recc - 0.5)))/(1.25^(log2(M)))); 
lif = (5*(Psi^2))+(2*Psi)+1 ;
Lb = 10*log10(lif);
end

\end{lstlisting}

\newpage

\end{document}